SAOS and LAOS rheology for differentiating chemical and physical crosslinking: A case study on PVA hydrogels

David Kogan and Moshe Gottlieb[a)]

*Department of Chemical Engineering, Ben Gurion University of the Negev, Beer Sheva, Israel*

**Abstract**

In this work, we have studied the viscoelastic behavior of chemically and physically crosslinked Poly(vinyl alcohol) (PVA) hydrogels near the critical gel point (GP) as well as further away from it, by means of small amplitude (SAOS) and large amplitude (LAOS) oscillatory shear experiments. Chemical crosslinking involved covalent bonding by means of glutaraldehyde as a crosslinker, while physical crosslinking was induced by freeze-thaw cycles. SAOS data analysis allowed evaluation of critical parameters such as the critical relaxation exponent $n$, gel strength $S$, and equilibrium modulus $Ge$, based on the dynamic self-similarity and fractal network structures at the GP. LAOS rheological data analysis showed that the chemically crosslinked system exhibited moderate strain-dependance due to the permanent covalent bonds, whereas the physically crosslinked system displayed significant strain-dependent nonlinearity due to strain dependent interactions at the crosslink entities. LAOS experiments, supported by Chebyshev coefficients and Lissajous-Bowditch plots, highlighted pronounced differences in nonlinear responses, underscoring the influence of crosslinking mechanisms on the network rheological behavior. The findings establish LAOS as a powerful tool for differentiating polymeric network structures, providing insights beyond those attained by conventional linear rheology.

---

[a)] Corresponding author; electronic mail: mosheg@bgu.ac.il

## Introduction

Viscoelastic materials may undergo a transition from a viscoelastic liquid (*sol*) to a viscoelastic solid (*gel*). This gelation process is the outcome of a percolative process (de Gennes 1979) initiated by the formation of clusters due to the induced interactions between individual molecules. At the transition point, known as the critical gel point (GP) separate clusters percolate to form an "infinite" cluster whose size encompasses the entire system. At this critical sol-gel transition point, the system is characterized by a unique state consisting of a newly formed, space-spanning three-dimensional percolated network coexisting with smaller, finite-sized clusters. Upon approaching the GP from the sol phase, the GP is characterized by the divergence of $\eta_0$, the zero-shear viscosity, and $M_w$, the weight average molecular weight. Crossing the GP and moving away from it into the solid gel phase leads to the emergence of non-zero equilibrium elastic modulus, $G_e$, and the insolubility of a fraction of the material, the insoluble gel. The percolative phase transition defined by the GP is a critical phenomenon with universal critical exponents. In the vicinity of the GP on either side, both the viscosity and the elastic modulus are characterized by a power-law dependence on the relative distance from the GP expressed in terms of some measure of the degree of crosslinking parameter, *P*, as follows (Suman and Joshi 2020):

$$\eta_0 \sim |\Delta P|^{-s} \quad \text{and} \quad G_0 \sim |\Delta P|^z \tag{1}$$

Here, $\Delta P$ is defined as $\Delta P = (P - Pc)/P_C$, i.e., the relative distance from the GP. $P_C$ is the value of $P$ at the critical transition point.

Winter and Chambon (W-C) (Chambon and Winter 1985) argued that at the GP both components of the shear modulus obtained by small amplitude oscillatory shear (SAOS) experiments, $G'$ the linear elastic modulus and $G''$ the linear viscous modulus, exhibit the same power law dependence with respect to the angular frequency, $\omega$, given by

$$G_C^{'} = \frac{G_C^{"}}{\tan(n\pi / 2)} = S\Gamma(1-n)\cos(n\pi / 2)\omega^n \tag{2}$$

Here, **n** is the critical relaxation exponent, **S** is the gel network strength, $\Gamma$ is the Euler gamma function, and the subscript $c$ denotes the value at the critical point. The relaxation exponent $n$ is restricted to values between 0 and 1. The case of $n$=0 corresponds to the limiting behavior of a Hookean solid. Thus, low values of $n$ correspond to a "stiffer" gel structure at the GP associated with higher values of $S$ and vice versa. Before the GP, the loss tangent, $tan\ \delta$= G"/G', decreases with increasing frequency, while past the GP, $tan\ \delta$ increases with increasing frequency. At the GP, $tan\ \delta$ becomes independent of the imposed frequency, suggesting it is independent of the probing time scale, and the two moduli are congruent. The power law rheology in the immediate vicinity of the GP results from a microstructure that is fractal in nature. The resulting self-similar morphology over a wide range of length scales promotes hierarchical relaxation timescales. Thus, the critical relaxation exponent $n$, is related to the two critical exponents $z$ and $s$ defined in Eq. 1, by the hyperscaling relation (Scanlan and Winter 1991):

$$n = \frac{z}{z+s} \tag{3}$$

The above-mentioned scaling relations have been verified by numerous authors for a large number of chemically and physically crosslinked polymeric systems such as poly dimethyl siloxane (Adam et al. 1997), polyvinyl alcohol (Kjøniksen and Nyström 1996; Joshi et al. 2020), polybutadiene (De Rosa and Winter 1994), epoxy (Adolf et al., 1990), polyurethane (Chambon et al. 1986), poly (ethylene oxide) (PEO), (Muller et al. 1991), cellulose (Lue and Zhang 2008), alginate (Zhao et al. 2016), and κ–carrageenan (Liu et al. 2015) to name a few. It was demonstrated that chemical and physical crosslinks are characterized by the same range of $n$ values, and systems with similar physical crosslinking mechanisms show a diversity of $n$ values. The relaxation exponent depends strongly on details such as the molecular weight of the precursor, stoichiometric ratio between the reactive components, and the amount of inert diluent in the material, but not on the nature of the crosslinking process and the size and structure of the crosslink point or zone (Scanlan and Winter 1991).

It is evident from the large amount of literature available that by employing linear rheological measurements and especially SAOS, one can determine $P_c$, the critical value of the degree of crosslinking at the GP, and the associated critical exponents. In addition, limited information regarding the structural properties of the crosslinked systems such as the fractal dimension $d_f$ of the percolated network (Muthukumar 1989), and the somewhat ambiguous gel strength parameter $S$ (Eq. 2), may also be obtained. Yet, details concerning the crosslinking process and the type of junction points or junction zones being formed cannot be determined from linear rheology.

Various materials exhibiting similar mechanical properties in the linear regime can display significantly different mechanical behavior in the non-linear regime. Therefore, the response to Large Amplitude Oscillatory Shear (LAOS) may yield information not accessible through linear measurements. For example, it was shown that LAOS could differentiate between suspensions of rigid and soft particles that have identical mechanical properties in the linear regime (Rogers 2018). Suman et al. (2023) showed that LAOS studies on colloidal gels (LAPONITE) near the GP provide microstructural insight unattainable by SAOS. Schlatter et al. (2005) demonstrated that using LAOS they were able to distinguish between linear and branched polyethylene melts. Goudoulas and Germann (2017) studied the differences between freshly prepared and aged crosslinked alginate-gelatin hydrogels. In the linear regime the mechanical properties of the fresh and aged gels were approximately the same while large differences were observed between the two hydrogels in the non-linear regime. Ramya et al. (2018) reported very large differences in the LAOS behavior between chemically crosslinked polyvinyl alcohol (PVA)-hyaluronic acid hydrogels and PVA hydrogels physically crosslinked by means of transient Borax crosslinks. Their study showed that the physical transient crosslinked system demonstrated LAOS strain stiffening independent of the degree of transient crosslinking or Borax concentration. In contrast, for the chemically crosslinked system non-linear behavior was not observed at all at low crosslinker concentrations over the entire range of strains examined. But an overshoot in G" and strain softening in G' were observed at high degrees of crosslinking and high strains. Although as pointed out above, there is only a limited amount of research available on the relationship between the type of crosslinking and the rheological properties of crosslinked polymeric systems under large deformations and even less so in the immediate vicinity of the GP, there are strong indications that non-linear rheology may provide the tools to differentiate between different types of network-forming mechanisms.

PVA hydrogels may be readily formed by several methods of chemical and physical crosslinking yielding mechanically stable hydrogels. Covalent chemical crosslinking is achieved by vulcanization reaction between the hydroxyl groups along the polymer backbone and one of several appropriate crosslinkers [e.g., glutaraldehyde (Figueiredo et al. 2009), or bis (β-hydroxyethyl) sulfone (Kumeta et al. 2004)]. Physical crosslinking of PVA may be achieved by freeze/thaw (f/t) cycles of PVA aqueous solutions (Hassan and Peppas 2000). Upon cooling below the Upper Critical Solution Temperature (UCST), microphase separated domains are formed serving as crosslinks and the hydrogel is held together either by these microdomains or, given enough time at the freezing temperature, by the substantially stronger PVA crystallites. Considerably weaker transient network PVA hydrogels are obtained by the complexation with Borax (Ewoldt and Bharadwaj 2013; Huang et al. 2017; Martinetti et al. 2018; Shim et al. 2023; Ramlawi et al. 2024). Hence, PVA hydrogels are well suited to examine the hypothesis that LAOS may provide the means to differentiate between different types of crosslinking methods. For this reason, we have selected to examine the gelation of two types of PVA hydrogels: a chemically vulcanized crosslinked system and a physically (f/t) crosslinked system. In the following sections, we show that the two types of systems while displaying similar rheological behavior in the LAOS regime, differ markedly in the LAOS regime.

**Materials and experimental section**

Poly (vinyl alcohol) (PVA) (Sigma Aldrich, China) 99% hydrolyzed with a weight average molecular weight of Mw=36 kDa and polydispersity of 1.5 (determined by SEC Postova model SC2000) was used as received. Glutaraldehyde (GA) (50% aqueous, Sigma Aldrich, Germany) served as crosslinker. Aqueous 0.5M hydrogen chloride (HCl) solution (Bio-Lab, Israel) was used as catalyst.

Stock aqueous solutions of PVA were prepared by adding the required amount of polymer (10% or 12% w/w) to de-ionized Millipore Direct-Q purified water (resistivity of 18.2MΩ-cm) in a round bottom flask fitted with a reflux condenser and CaCl$_2$ trap. The mixture was heated using an oil bath to 90-100 ℃ under vigorous stirring for four hours until a clear solution devoid of aggregates was obtained. The solution was allowed to cool to room temperature.

**Chemical (vulcanization) crosslinking (CX).**

The amount of aq. solution of the crosslinker required to achieve the desired degree of crosslinking was added to the 10% aq. PVA solution in a 20 mL vial, followed by pipetting HCl solution to achieve pH of 2.5. The mixture was vigorously mixed in the closed vial. The reaction mixtures were subsequently poured into petri dishes sealed with parafilm to prevent water evaporation and kept for two days at room temperature to allow complete curing. The composition of the systems prepared are listed in Table 1 in terms of SR, the stoichiometric ratio between the moles of PVA monomers and the moles of GA in the mixture. One end of the glutaraldehyde molecule reacts covalently with two adjacent hydroxyl groups on the PVA backbone, while the other end reacts with the hydroxyl groups of a neighboring polymer molecule, resulting in a four functional crosslink site. Although in principle, intra-molecular crosslinking is possible, we expect its effect to be negligible due to the relatively high PVA concentration (well above the overlap concentration $C^*$). As can be observed in Table 1 the amount of glutaraldehyde molecules is considerably smaller than the number of hydroxyl groups along the PVA backbone and as result it is expected to be completely consumed during the vulcanization reaction. The degree of crosslinking, $X$, defined as the weight average number of crosslinks per polymer chain (listed in Table 1), is obtained by assuming that each GA molecule in the system forms one crosslink by engaging two PVA monomers (molar mass $M_0$) per chain. Thus, $X = SR / (Mw\ PVA/2*M_0)$. The relative distance from the critical gel point $\Delta P = (X/X_c - 1)$, is listed as well in Table 1. The value of $X_c$ the critical degree of crosslinking at the GP, has been obtained as discussed in the results section in conjunction with Fig 1.

Table 1 Compositions of the fully reacted PVA-GA samples

| SR (mol PVA monomer)/ (mol GA) | X (crosslinks/chain) | $\Delta P$ ($X/X_c$-1) |
|---|---|---|
| 498 | 0.82 | -0.18 |
| 450 | 0.91 | -0.10 |
| 437 | 0.94 | -0.06 |
| 425 | 0.97 | -0.03 |
| 412 | 1.00 | 0 |
| 400 | 1.03 | 0.03 |
| 387 | 1.06 | 0.06 |
| 372 | 1.10 | 0.1 |
| 350 | 1.15 | 0.15 |
| 300 | 1.36 | 0.36 |

**Physical crosslinking (PX).**

Physically crosslinked PVA hydrogels were prepared by the inverse-quenching freeze-thaw (f/t) technique (Hassan and Peppas 2000; Avallone et al., 2021). The desired amounts of 12% (w/w) PVA solutions were poured into petri dishes, closed, and sealed with parafilm. The solutions initially at $T_i = 25℃$, were frozen by cooling to $T_w = -20℃$ (well below the UCST) for a period $t_w$ (from 1h up to 24h). At the end of the freezing stage, the samples were allowed to thaw at $T_f = 25℃$ for 24 h, completing one f/t cycle. Some samples were subjected to repeated f/t cycles with $t_w = 24hr$ for all cycles. Physical crosslinks are formed by an initial microphase separation process followed by crystallization of PVA segments within the microphase separated domains (Ricciardi et al. 2004a and b; Holloway et al. 2013). Since the melting point of PVA crystallites is considerably above room temperature (between 50-70$^0$C, Ricciardi et al. 2004b), the hydrogels formed by the f/t method employed here and kept at room temperature, remain stable for extended periods of time (months) without any detectable change to their properties. Polymer crystallization kinetics are slow even within the confinement of the microphase-separation domains. If $t_w$ is not sufficiently long to allow crystallites formation, the physical hydrogel network will be held together only by the microphase separation domains. In this case the hydrogel will disintegrate when the system is thawed above its redissolution temperature, which is considerably lower than the melting temperature. Such a breakup of PVA hydrogels has been observed at 7.7$^0$C (Suman and Joshi 2020), and at 20$^0$C (Joshi et al. 2020).

**Rheological measurements.**

The measurements were carried out interchangeably on a TA-ARES strain-controlled rheometer, and an Anton Paar MCR702 dual-motor rheometer employed as a strain-controlled rheometer (separated motor-transducer Merger and Wilhelm 2014). Either 25 mm or 50 mm sandblasted parallel plate geometries and an advanced Peltier system for temperature control were used. A water-saturated enclosure was mounted on the geometry to prevent water evaporation when testing water swollen hydrogels. The temperature was kept at 25℃. Repeat measurements carried out on both rheometers yielded similar results. To determine the equilibrium modulus, *Ge*, stress relaxation experiments were conducted. The sample was subjected to a constant strain (in the linear regime), and the shear stress was recorded as a function of time for up to 200 seconds. Using oscillatory SAOS experiments the critical gel point was determined. Frequency sweeps were performed in the range of 0.1-100 rad/s at a strain value well within the linear regime.

The critical gel point (GP) was determined by the W-C method (Eq. 2) i.e., identifying the degree of crosslinking at which $G'$ and $G''$ are congruent over the entire examined frequency range, and as a result, tan$\delta$ is independent of frequency. Next, the LAOS-strain rheological behavior was determined by subjecting the samples to a strain sweep at several selected frequencies within the range of 0.1-100 rad/s but, often only between 1 and 10 rad/s. The sample was placed between the parallel plates and a strain amplitude sweep was carried out at increasing strains followed by decreasing strains, to ascertain sample integrity during the strain sweep. Fourier transform, and Lissajous plots were obtained using RheoCompass software provided by Anton Parr or MITlaos (Ewoldt et al. 2008a) for the experimental data obtained by the ARES rheometer.

## Results and discussion

### SAOS

The critical GP was determined by conducting at least two decades of angular frequency SAOS tests on both CX and PX PVA systems (strain amplitudes varied between 0.1-20%, all within the linear viscoelastic regime) and employing the W-C analysis (Chambon et al. 1986). Figure 1 illustrates the dependence of *tan$\delta$* on the degree of crosslinking parameter, *X*, and the relative distance from the critical GP, $\Delta P = (X/Xc - 1)$ for the CX system and on the freezing time, $t_w$, for the PX system. Each curve corresponds to a different oscillation frequency value. For both systems, it can be observed that in the pre-gel state, *tan$\delta$* values decrease with increasing $\omega$ while in the post-gel state, the trend is reversed. The intersection of the different $\omega$ curves marks the transition from a viscoelastic liquid to a viscoelastic solid, since *tan$\delta$* is independent of ω due to self-similarity phenomena at this unique point. The values of the critical GP obtained from Figure 1 are *$X_c$ = 1.00* for CX system and *$t_{w,c}$=6.3h* for the PX system. The rheological fingerprint at the critical GP is the identical dependence of both moduli on $\omega$, $G' \sim G'' \sim \omega^n$, following Eq. 2. Based on the latter, the critical relaxation exponent, *n*, was found to be 0.70 and 0.82 for the CX and PX systems, respectively. The gel strength, *S*, was also determined and found to be 1.8 Pa·s$^n$ for CX and 0.1 Pa·s$^n$ for PX. The critical gel is generally soft and fragile when $n \rightarrow 1$; thus, *S* is small. While the value of *n* for PX is only slightly higher than the one for CX, the *S* value is an order of magnitude smaller revealing that the CX hydrogel is considerably stiffer than the PX one. Joshi and coworkers (Joshi et al. 2020; Suman et al. 2020; Bhattacharyya et al. 2023) using several different f/t protocols to generate physically crosslinked PVA obtained n=0.76-0.85 and S=0.1 Pa·s$^n$. These values are in agreement with the values obtained here despite the fact these hydrogels were formed only by microphase separation with no crystallization, as evident from their dissolution at various temperatures below room temperature.

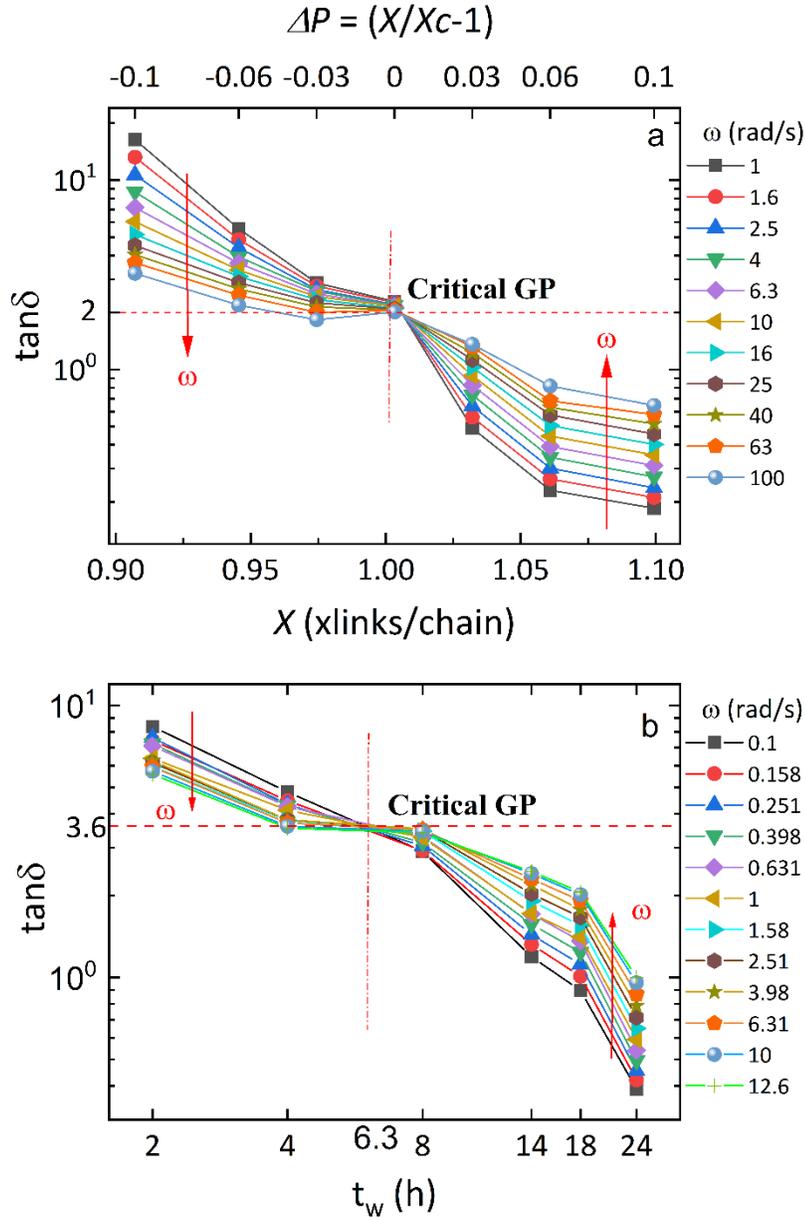

Figure 1 The phase shift tan$\delta$ for different angular frequencies, $\omega$, (a) as function of the degree of crosslinking parameter, X, and the relative distance from the GP, $\Delta P=(X/X_c-1)$, for the CX system and (b) as function of the freezing waiting time, $t_w$ for the PX system. The red arrows represent the direction of increasing values of $\omega$. The dashed lines indicate the position of the critical GP. Strain amplitudes varied between 0.1-20% for the different samples.

Frequency sweep experiments were utilized to estimate the zero-shear viscosity, $\eta_0$, for the different gelling systems before the GP. This was achieved by fitting $|\eta*|$ data to the Cross model and extrapolation to zero frequency $\eta_0 = \lim_{\omega \to 0} |\eta*|$. Stress relaxation experiments after the GP were utilized to determine the equilibrium modulus $Ge$ by extrapolating the values of $G(t)$ to long times $G_e = \lim_{t \to \infty} G(t)$ (Winter et al. 1994), as illustrated in Figure 15 in Appendix 1. Since the viscosity curves for the PX system showed little tendency towards a plateau at low frequencies (Fig. 3b), we resorted to an iterative process to obtain a rough estimate for these values. Details of the procedure are provided in Appendix 2 below. The values of $\eta_0$ and $G_e$ for the CX systems examined here are tabulated in Table 2 and for the PX systems in Table 3.

Table 2 Zero shear viscosity, $\eta_0$, equilibrium modulus, $Ge$, and the longest relaxation time, $\tau_{max}$ for the chemically crosslinked systems. ($\tau_{max}$ has been obtained as described below in conjunction with Fig. 3).

| $X$ (Crosslinks/Chain) | $\Delta P = (X/X_c - 1)$ | $\eta_0 \, (Pa \cdot s)$ | $Ge \, (Pa)$ | $\tau_{max} \, (s)$ |
|---|---|---|---|---|
| 0.82 | -0.18 | 0.22 | - | $6.1 \cdot 10^{-4}$ |
| 0.91 | -0.10 | 0.50 | | $1.3 \cdot 10^{-2}$ |
| 0.94 | -0.06 | 2.04 | - | 2.4 |
| 0.97 | -0.03 | 3.20 | - | 12.7 |
| 1.00 | 0.00 | $\to \infty$ | $\to 0$ | $\to \infty$ |
| 1.03 | 0.03 | - | 4 | 0.35 |
| 1.06 | 0.06 | - | 25 | $2.4 \cdot 10^{-2}$ |
| 1.10 | 0.10 | - | 50 | $9.1 \cdot 10^{-3}$ |
| 1.15 | 0.15 | - | 150 | $2.0 \cdot 10^{-3}$ |
| 1.36 | 0.36 | - | 812 | $1.6 \cdot 10^{-4}$ |

Table 3 Zero shear viscosity, $\eta_0$, equilibrium modulus, $Ge$, and the longest relaxation time, $\tau_{max}$ for the physically crosslinked system. * The system at $t_w$=8h is not strictly at the GP but somewhat past it, as observed in Fig. 1b. Since G' and G" are almost congruent for this system (slopes of 0.89 and 0.82 respectively) it was used as the reference line for the determination of the longest relaxation times in Fig. 3b.

| $t_w\ (h)$ | $\eta_0\ (Pa{\cdot}s)$ | $Ge\ (Pa)$ | $\tau_{max}\ (s)$ |
|---|---|---|---|
| 2 | 0.45 | - | 2.6 |
| 4 | 0.80 | - | 65.5 |
| 8* | $\rightarrow\infty$ | $\rightarrow 0$ | $\rightarrow\infty$ |
| 14 | | 0.025 | 22 |
| 18 | | 0.06 | 9.44 |
| 24 | | 1.00 | 0.31 |

The experimental values of the zero-shear viscosity and the equilibrium modulus for CX are plotted in Figure 2 as function of the relative distance from the GP and fitted to the scaling laws defined in Eq. 1 (de Gennes 1979; Adam et al. 1997, Suman and Joshi 2020). The values $s$=1.07 ±0.36 ($R^2$=0.84) and $z$=2.63±0.60 ($R^2$=0.97) are obtained. Inserting these values into the hyperscaling relation Eq. 3, the critical relaxation exponent is recovered n=0.71±0.02, in excellent agreement with $n$=0.70 determined by the W-C analysis. As pointed out by Adam et al. 1997 and Suman and Joshi 2020 the values of $s$ and $z$ obtained for different systems and by different authors are widely scattered. The values obtained here are, within the error margins, in agreement with those predicted by the Rouse model ($s$=1.33 and $z$=2.67) and in agreement with several of the chemically crosslinked systems reported in Table I of Suman and Joshi (2020).

Similar analysis was carried out only for the gel side of the PX systems due to the lack of sufficient number of $\eta_0$ experimental values (only two systems were available below the gel point due to the relatively low and hardly measurable, values of the moduli of physically crosslinked PVA solutions below the gel point). The value of $z$=4.0±1.2 was obtained from the $G_e$ data in agreement with the value reported for physically crosslinked cellulose and within the error margins for different PVA systems (Table I Suman and Joshi 2020).

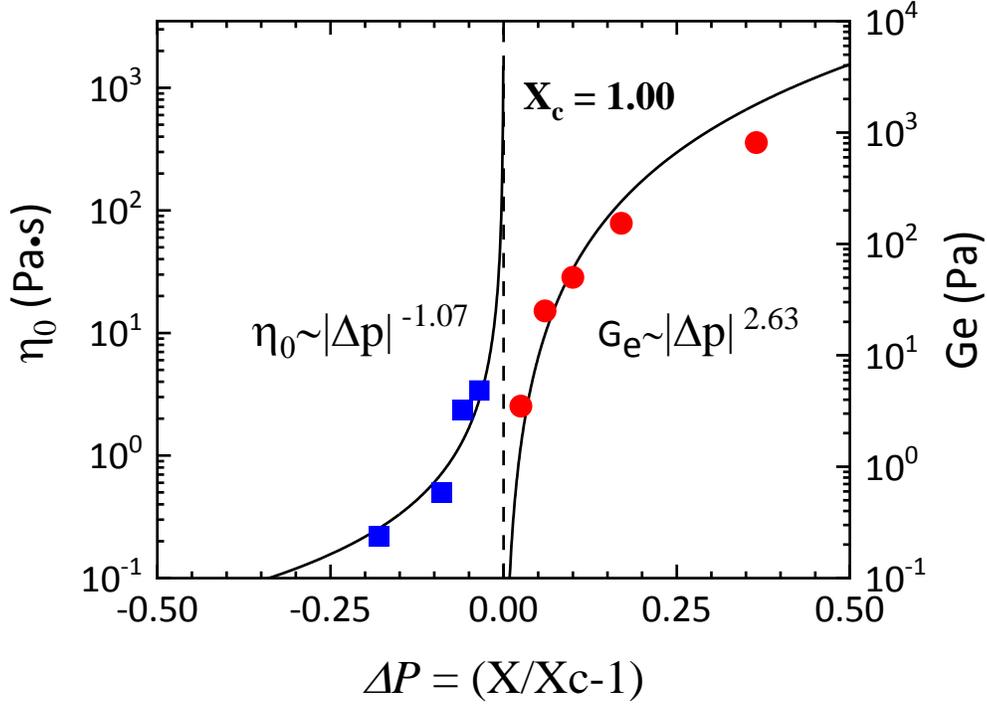

Figure 2 Evolution of $\eta_0$ and $G_e$ as a function of $\Delta P$ for chemically crosslinked PVA. The critical GP is represented by the dashed line. The solid lines represent the power-law fitting for $\eta_0$ and $G_e$ to Eq.1.

The values of the magnitude of the complex viscosity, $|\eta^*|$, as a function of $\omega$, for different CX systems either before or after the GP are depicted in Figure 3a and for PX in Figure 3b. Each set of points represents data collected at one specific distance from the GP. Approaching the GP from below, the values of the complex viscosity increase gradually due to cluster formation. Past the GP, the complex viscosity further increases due to the buildup of the gel. The critical GP system serves as the reference line for describing the evolution of the rheological properties and for distinguishing between the sol and the gel phases.

The power-law behavior at the critical GP state is an expression of mechanical self-similarity (fractal behavior), and it is formulated as follows (Suman and Joshi 2020):

$$\left|\eta^*\left(\omega,\Delta P\right)\right| = aS\omega^{n-1} \qquad (4)$$

Where $a = \pi/\left(\Gamma\left(n\right)\sin\left(n\pi\right)\right)$. Thus, with $n$=0.7 from the W-C analysis or 0.71±0.02 from the scaling analysis, it is expected according to Eq. 4 that $|\eta^*|\sim\omega^{0.29\pm0.02}$. The experimental data corresponding to the critical gel (green open triangles and solid line in Figure 3a) yields a straight line (the GP line) with a slope of -0.27 in complete agreement with the value predicted above. In a similar fashion the slope of the GP line for the PX system (green open triangles and solid line in Figure 3b) is -0.18 in complete agreement with n=0.82 obtained by the W-C analysis described above. Although the system after 8h of freezing is somewhat past the GP as observed in Fig. 1b, the close similarity in the slope values of G' and G'' as function of frequency and the agreement between the slope of the gel point line in Fig. 3b and the $n$ value obtained by the W-C method are indications that selection of this system to represent the critical GP is an acceptable approximation.

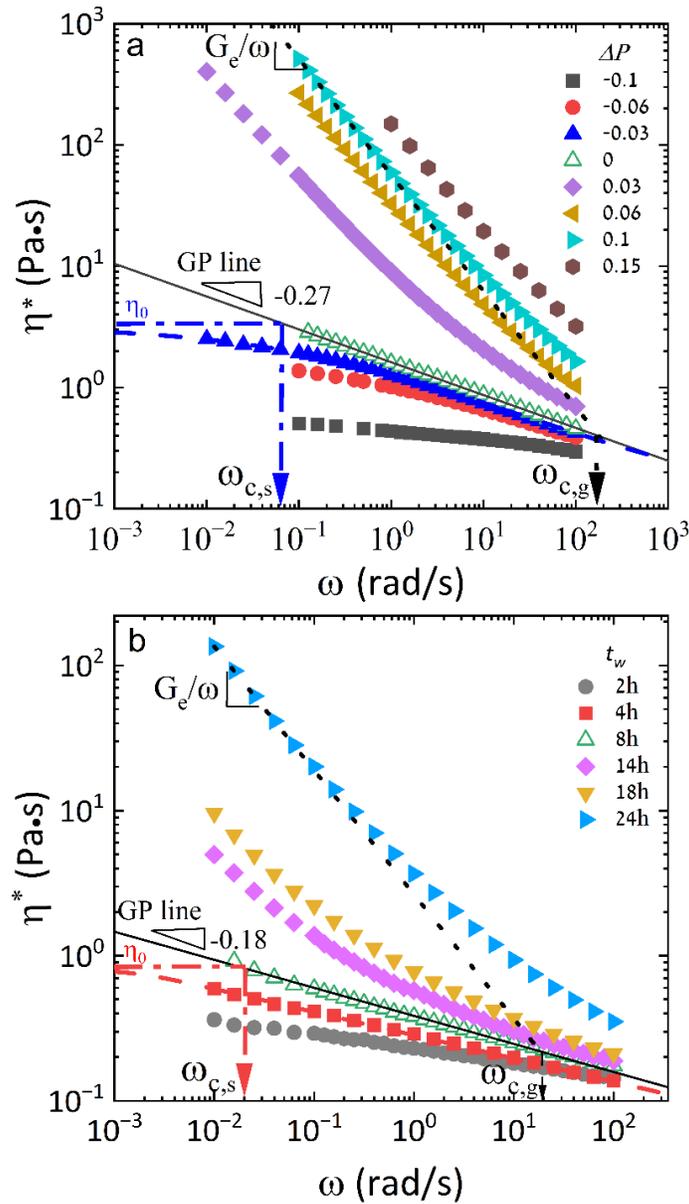

Figure 3 The magnitude of the complex viscosity as function of angular frequency for (a) CX and (b) PX PVA. Each curve represents a system at a different relative distance from the critical GP. The solid line represents the power-law fitting to the data at the critical GP (n-1= -0.27 for CX and -0.18 for PX). The extrapolation to obtain the zero-shear viscosity in the pre-gel states is shown by the dashed-lines. The dotted lines represent the low $\omega$ fit to the expression $|\eta^*|=Ge/\omega$ for the post-gel states. The critical frequencies, $\omega_{c,s}$ and $\omega_{c,g}$ are obtained from the intersection of the asymptotes with the GP line for the pre- and post-gel states respectively as demonstrated by the dash-dotted arrows. The inverse of these critical frequencies defines the longest relaxation times in the pre- and post-gel states.

On both sides of the GP the values $|\eta^*|$ increase with decreasing frequency. In the pre-gel state, these values level off approaching asymptotically the zero-shear viscosity at vanishing frequencies. In the post gel state (data above the GP line in Figure 3) the complex viscosity values increase sharply with decreasing frequency towards the asymptote $|\eta^*|\sim\omega^{-1}$ characteristic of a solid like behavior leading to the relation $|\eta^*|=G_e/\omega$ at vanishing frequencies.

The longest relaxation time $\tau_{max}$ was estimated as follows: For each crosslinked system in the pre-gel state the point of intersection between the horizontal asymptote to $\eta_0$ with the GP line determines the critical angular frequency in the sol-phase, $\omega_{c,s}$ as demonstrated in Figure 3 for the case of $\Delta P$=-0.03 (Figure 3a) and $t_w$=4h (Figure 3b). The longest relaxation time is identified as the inverse of the critical frequency associated with this intersection point. In the gel phase the critical angular frequency $\omega_{c,g}$, was determined using the low $\omega$ values as well. The point of intersection between the low frequency asymptote $Ge/\omega$ with the GP line (e.g., the dotted line associated with the $\Delta P$=0.1 in Figure 3a and $t_w$=24h in Figure 3b) determines the critical frequency and its inverse is taken as the longest relaxation time. The values for the longest relaxation times determined by the procedure described above are listed in Tables 2 and 3.

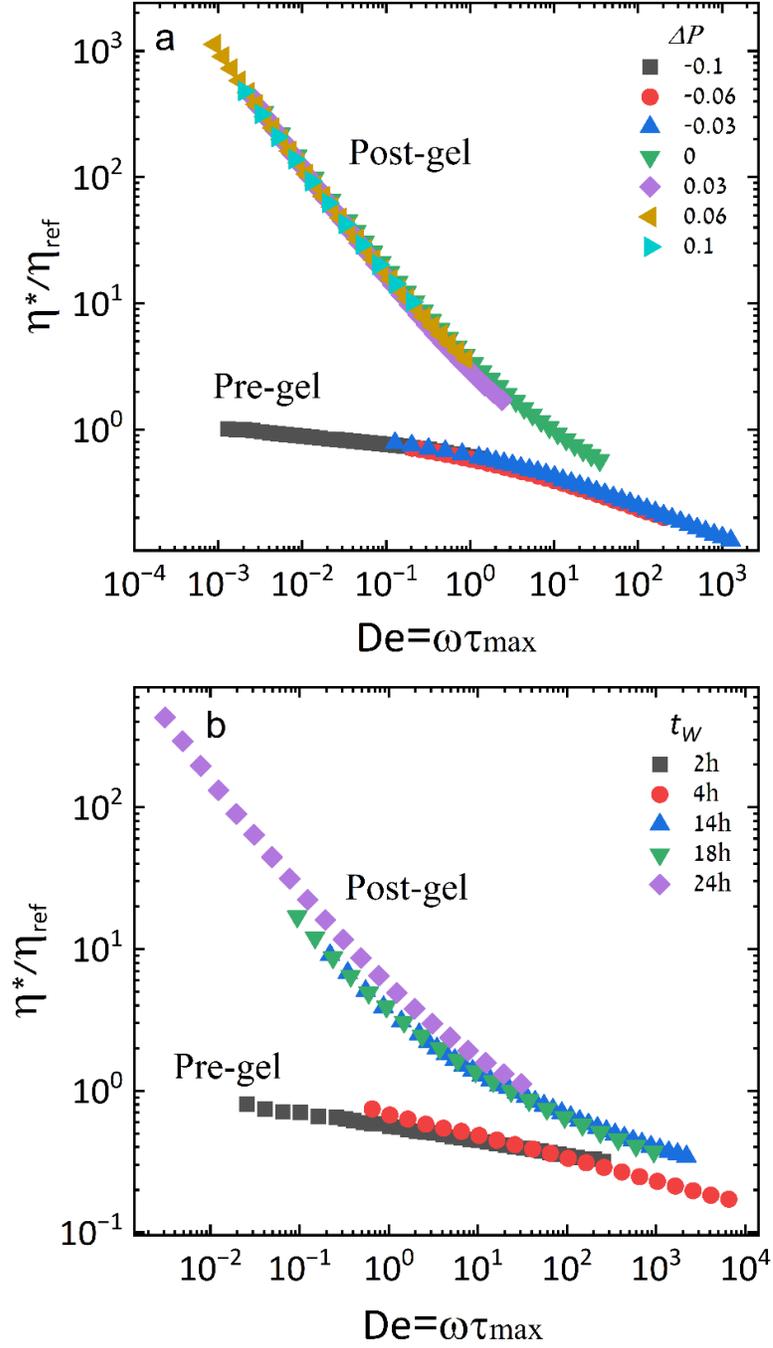

Figure 4 Normalized $\eta^*$ as a function of the dimensionless Deborah number De=$\omega\tau_{max}$ for (a) CX and (b) PX systems. In the pre-gel state $\eta^*$ is normalized by the zero-shear viscosity and in the post-gel state by $G_e\tau_{max}$.

The complex viscosity values are now normalized by a reference viscosity $\eta_{ref}$, where $\eta_0$ is used as the reference in the pre-gel state and $Ge\tau_{max}$ in the post- gel state.

Normalized viscosity values are plotted in Figure 4 as a function of the dimensionless Deborah number, $De = \omega\tau_{max}$. resulting in two master curves one for the sol phase and one for the gel phase, both spanning six orders of magnitude of $De$ numbers. The existence of these master curves is an indication that on either side of the critical GP, the longest relaxation times diverge in a similar fashion. Thus, the relaxation time distribution remains preserved.

## LAOS

When the amplitude of the applied sinusoidal oscillatory shear exceeds some characteristic critical value, the resulting material response departs from the sinusoidal pattern defining the onset of the non-linear viscoelastic regime. In this non-linear regime $G'$ and $G''$ are not sufficient to fully characterize the nature of the elastic and viscous non-linear response of the material. As result, several schemes have been developed over the years to analyze the response to LAOS and relate it to the material microstructure. The Fourier Transform (FT) rheological analysis (Wilhelm 2002; Merger and Wilhelm 2014; Giacomin et al. 2015; Merger et al. 2016) is based on the examination of the non-linear harmonics, i.e., the elastic and viscous non-linear shear moduli fundamental harmonics $G_1'$ and $G_1''$, and the subsequent higher non-linear harmonics $G_j'$ and $G_j''$ (odd j≥3). While the fundamental harmonics are related to the global dynamic response of the material the third harmonics has been associated with local microstructural dynamics. For example, Double Quantum Proton NMR which has been shown to reveal local dynamics (Saalwächter 2007, Saalwächter et al. 2007, Valentin et al. 2010), was shown by Nie et al. (2019) to relate the third harmonic intensity to local dynamics in natural rubber. The third harmonic was also shown to relate to the local particle-polymer interactions in filled rubbers (Schwab et al. 2016). Further refinement has been suggested by means of the Chebyshev polynomial fit (Ewoldt et al. 2008b; Ewoldt and Bharadwaj 2013; Bharadwaj et al. 2017). which has been claimed to provide additional interpretability to the microstructural dynamics based on the sign of the third harmonic. Yet as pointed out by Shim et al. (2023), a breakdown in the interpretability is expected at higher strains when contributions from the larger harmonics become significant. In addition, the reliance on complete oscillatory periods precludes the use of transient data. An alternative approach termed the sequence of physical processes (SPP), provides information concerning the instantaneous local elastic and viscous response derived from transient data (Rogers and Lettinga 2012; Rogers 2017; Choi et al. 2019). The main advantage of the SPP approach over the Chebyshev fit relates to yielding materials. While all the above-mentioned methods are based on different mathematical analyses of the same experimental technique, recently a new experimental termed "recovery rheology" has been developed (Kamani et al. 2023). Although recovery rheology has been shown to enable deeper understanding of the LAOS response and resolve the differences between the different mathematical approaches (Shim et al. 2023 and references therein) these benefits are afforded at the expense of considerably larger experimental effort.

### Strain sweeps of chemically crosslinked systems

Four PVA CX systems at equal distances on either side of the GP ($\Delta P = \pm 0.1$, and $\pm 0.03$) were subjected to LAOS strain amplitude sweeps at several frequencies spanning the range of 1-10 rad/s. The first harmonic nonlinear moduli, $G_1^{'}$ and $G_1^{''}$, normalized by the linear moduli values are depicted in Figure 5. The frequency values are expressed in terms of the $De$ number. $De = \omega \tau_{max}$ to allow comparison between the time (frequency) dependent properties of systems with largely different dynamics as indicated by the values of the longest relaxation times for the pertinent systems (cf. Table 2).

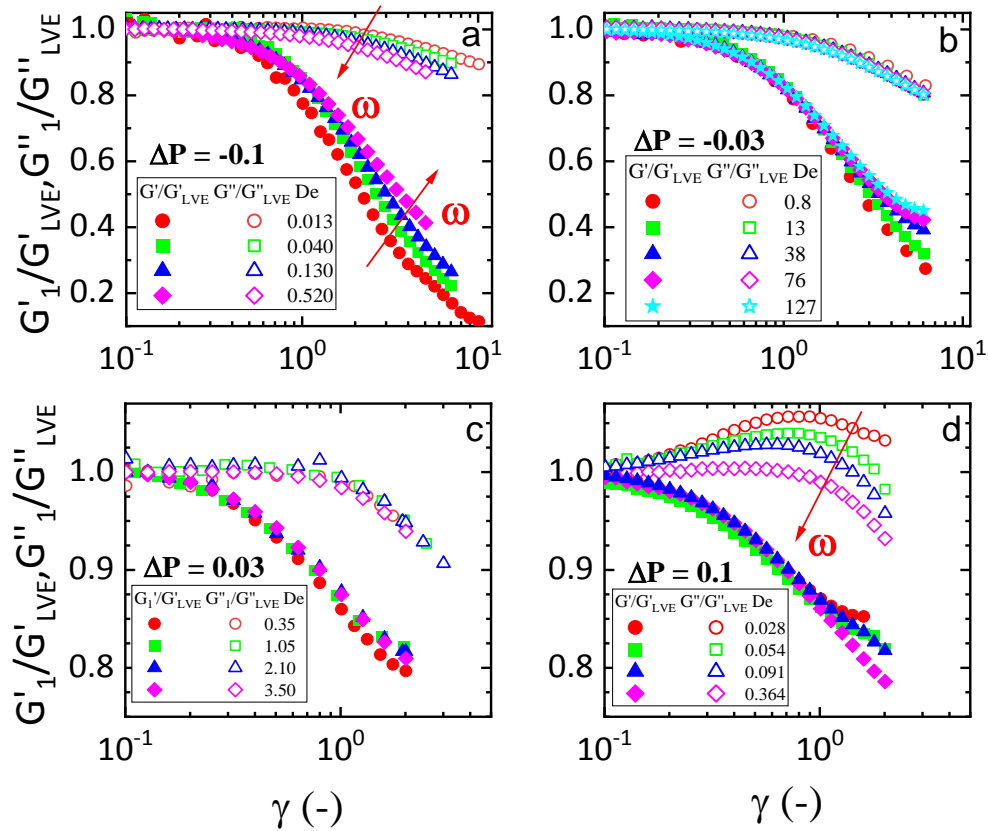

Figure 5 Strain dependence of the first harmonic non-linear moduli normalized by the corresponding modulus value obtained in the linar regime, for systems at different distance from the GP. (a) ΔP = -0.1. (b) ΔP = -0.03 (c) ΔP = +0.03 (d) ΔP = +0.1. The red arrows indicate the direction of increase in frequency.

For all four samples the effect of large strains is manifested by a substantial decrease in the values of the first harmonic storage modulus $G_1^{'}$ and a considerably smaller, yet discernible drop in $G_1^{''}$ values. The notable exception is the case for $\Delta P = +0.1$ which displays an initial increase and overshoot in $G_1^{''}$ before the decline. Another salient feature is the considerably larger non-linear effect in the pre-gel systems in comparison to the modest effect in the post-gel systems. This difference is probably due to the higher resistance to large deformations of the crosslinked post-gel network relative to the resistance offered by the system prior to the three-dimensional network formation.

The high strain amplitudes impose alignment of the long polymeric chains/clusters with the deformation field resulting in decreased mechanical properties manifested by the strain-thinning phenomena – the decrease in both moduli values (Hyun et al. 2002; Sim et al. 2003). At $\Delta P = -0.1$ increase in $\omega$, results in contradicting effects on the two components of the modulus. As the frequency increases a stronger decrease in the first harmonic loss modulus $G_1^{''}$ is observed, i.e., a stronger non-linear effect. In contrast, a weaker decrease in the values of $G_1^{'}$ is observed as the frequency increases, i.e., the magnitude of the non-linear effects in the first harmonic storage modulus $G_1^{'}$ decrease with increasing frequency (c.f. arrows in Figure 5a). Closer to the critical GP ($\Delta P=-0.03$, Figure 5b), the frequency dependence vanishes, and no such dependence is observed over two orders of magnitude of Deborah number. Upon crossing the critical GP, a similar observation is made: at $\Delta P=0.03$, (Figure 5c), no dependence on frequency over the entire range of strains and frequencies tested, similar to what has been observed for $\Delta P=-0.03$. Moving further away from the GP into the gel regime ($\Delta P=0.1$, Figure 5d) frequency dependence is observed for $G_1^{''}$ similar to the observation for $\Delta P=-0.1$ (Figure 5a). Yet, while in the latter the nonlinear effect (decrease in moduli values) increases with increasing frequency, in the former the main non-linear effect – the overshoot in $G_1^{''}$ becomes smaller with increasing frequency. We suggest that the overshoot behavior arises from the fact that the entangled chains between the crosslinks and the large clusters that are not connected to the network initially resist the large deformations until the point in which they align with the flow field, followed by a drop in $G_1^{''}$ values.

An important observation from the strain sweep experiments discussed above is the independence of the moduli in the vicinity of the critical GP from the imposed frequency (over two orders of magnitude of the Deborah number). This result suggests that the self-similarity holds in the non-linear regime.

### W-C in the non-linear regime

To further investigate the assumption that self-similarity holds in the non-linear regime, the two gelling systems (CX and PX) were subjected to strain sweep experiments at different frequencies between 1 and 10 rad/s. The *tanδ* values at a relatively high strain amplitude of 200% are depicted in Figure 6. The results of the ensuing W-C analysis in the linear and non-linear regimes are tabulated in Table 4.

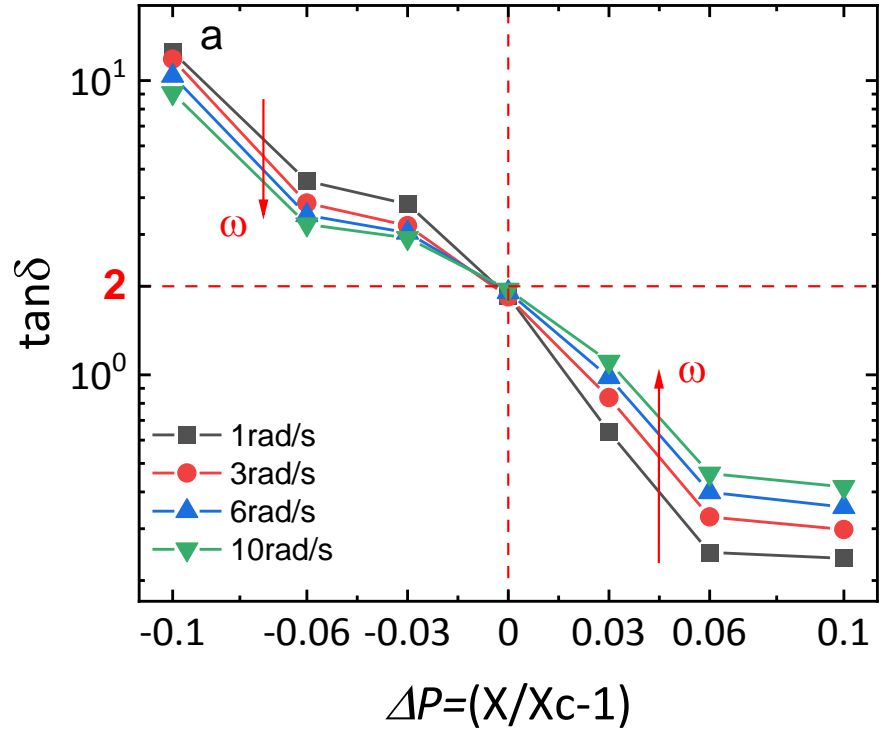

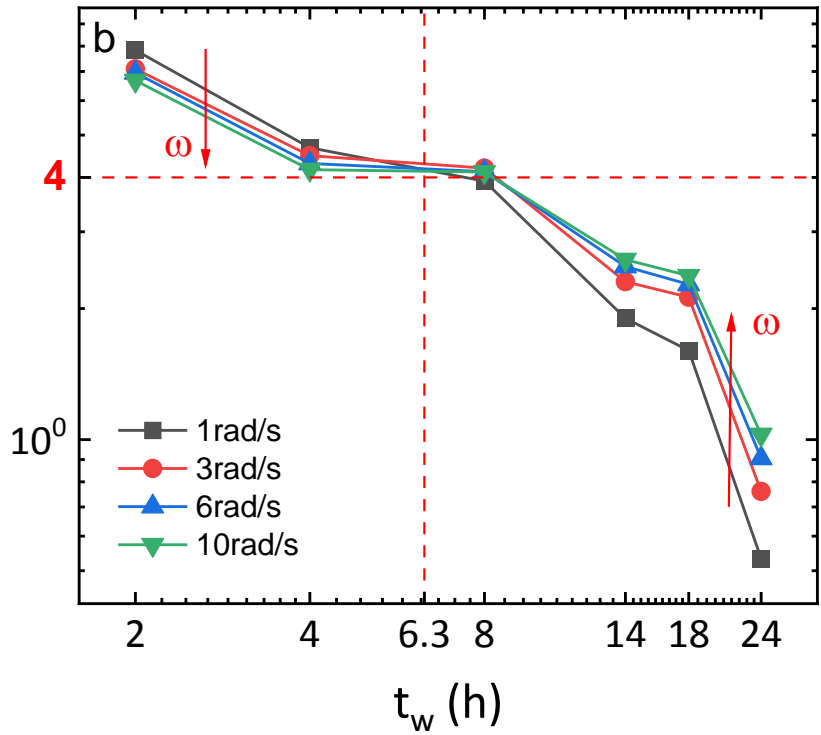

Figure 6 The loss tangent tan$\delta$ at different angular frequencies, $\omega$, and strain amplitude $\gamma_0$=200% (a) as function of the relative distance from the GP, $\Delta P$=(X/X$_c$-1), for the chemically crosslinked PVA system and (b) as function of the freezing time, t$_w$ for the physically crosslinked system. The dashed lines indicate the position of the critical GP.

Muthukumar (1989) developed a relationship between the critical relaxation exponent and the fractal dimension, $d_f$, of the percolated network at the critical GP:

$$n = \frac{d\left(d+2-2d_f\right)}{2\left(d+2-d_f\right)} \qquad (5)$$

Here $d$ is the dimension of the system, which in the PVA physical and chemically crosslinked systems is three. The expression considers the complete screening of the excluded volume interactions. The fractal dimensions for both systems as calculated using Eq. 5 are indicated as well in Table 4.

Table 4. W-C analysis in the linear and nonlinear regimes

|  | CX PVA | | PX PVA | |
|---|---|---|---|---|
|  | Linear | Non-Linear | Linear | Non-Linear |
| $P_c$ or $t_{w,c}$ | 1.0 | 1.0 | 6.3 | 6.3 |
| $tan\delta_c$ | 2.0 | 2.0 | 3.6 | 4.0 |
| $n$ | 0.7 | 0.7 | 0.82 | 0.84 |
| $S\ (Pa{\cdot}s^n)$ | 1.8 | 1.5 | 0.1 | 0.06 |
| $d_f$ | 1.74 | 1.74 | 1.56 | 1.53 |

The results presented in Table 4 clearly indicate that apart from the values for the gel strength S, the values obtained by the W-C analysis for the LAOS data are identical to those obtained by SAOS for both chemically and physically crosslinked networks. These results suggest that although the applied deformation is large, the critical properties at the GP are preserved as well as its topological structure as indicated by the fractal dimension. The results presented here indicate that the W-C analysis is not limited to the SAOS regime and may be used in the LAOS regime as well. The decrease in the gel strength parameter $S$ in the non-linear regime is a manifestation of the reduction in the network resistance to deformation at higher strains in agreement with the decline in moduli values depicted in Figure 5. Similar reduction in S values with increasing strains has been reported for a variety of other crosslinked systems (Kogan and Gottlieb 2025).

**Comparison between CX and PX systems in the vicinity of the gel point**

Investigation limited to the first non-linear harmonic moduli is insufficient to completely encompass the non-linear LAOS response of materials. To better characterize the nature of the elastic and viscous non-linear response, the third harmonics, and the corresponding Chebyshev polynomial fit (Ewoldt et al. 2008b; Ewoldt and Bharadwaj 2013) were employed. We have opted to use the FT and Chebyshev fit rather than the more elaborate SPP or recovery rheology methods described above, since even these simpler analysis methods provide the means to distinguish between the different types of crosslinks. The use of the Chebyshev method is also justified by the lack of any harmonics higher than the third harmonic and no transients are expected. The Chebyshev coefficients are directly related to the Fourier series of the resulting non-linear stress response, as follow (Bharadwaj and Ewoldt 2015)

$$\sigma(t;\omega,\gamma_0) = \gamma_0 \sum_{j=1:odd} \left\{ G_j^{'}(\omega,\gamma_0)\sin(j\omega t) + G_j^{''}(\omega,\gamma_0)\cos(j\omega t) \right\} \qquad (6)$$

where the Fourier moduli $G_j^{'}(\omega,\gamma_0)$ and $G_j^{''}(\omega,\gamma_0)$ are functions of ω and $\gamma_0$.

Decomposition of the resulting stress into viscous and elastic parts (Cho et al. 2005; Läuger and Stettin 2010; Ewoldt and Bharadwaj 2013; Natalia et al. 2020) directly relates the Fourier moduli coefficients to the Chebyshev coefficients as follows:

$$e_j = G_j^{'}(-1)^{(j-1)/2} \qquad (7)$$

$$v_j = G_j^{''} / \omega \qquad (8)$$

Here $e_j(\omega, \gamma_0)$ and $v_j(\omega, \gamma_0)$ are the elastic and the viscous non-linear Chebyshev coefficients, respectively. The first harmonic coefficients $e_1(\omega, \gamma_0) = G_1^{'}(\omega, \gamma_0)$ and $v_1(\omega, \gamma_0) = \eta_1^{'}(\omega, \gamma_0)$ are measures of average elasticity and average dissipation, respectively. The third harmonic Chebyshev coefficients $e_3(\omega, \gamma_0)$ and $v_3(\omega, \gamma_0)$ indicate local non-linear dynamic response. The advantage of using Chebyshev fitting is that the third non-linear Chebyshev coefficients provide physical meaning at large deformations in contrast to Fourier moduli coefficients. Positive values of $e_3$ indicate strain stiffening and negative values strain softening. In the same manner, $v_3 > 0$ is interpreted as shear rate thickening and $v_3 < 0$ as shear rate thinning (Ewoldt et al. 2008b; Ewoldt and Bharadwaj 2013; Martinetti et al. 2018; Singh et al. 2018). Yet, as pointed out (Poulos 2013, Shim 2023) the strain stiffening/softening interpretation may breakdown at higher strains when fifth and higher harmonics emerge.

Two PVA hydrogels both in the post-gel state, one chemically crosslinked ($\Delta P$=0.03) and the second physically crosslinked ($t_w$=24h), exhibit similar rheological characteristics in the linear regime, as illustrated in Figure 7. Their dynamical similarities are expressed in terms of the longest relaxation times ($\tau_{max}$ = 0.35 s and 0.31 s for CX and PX, respectively) and their linear complex moduli spectra. The identical frequency dependence of the linear moduli for these two systems is further demonstrated by the identity in *tan$\delta$* values over three decades of frequency as depicted in Figure 7. These two systems rheologically similar in the linear regime, were further examined by means of non-linear rheological measurements.

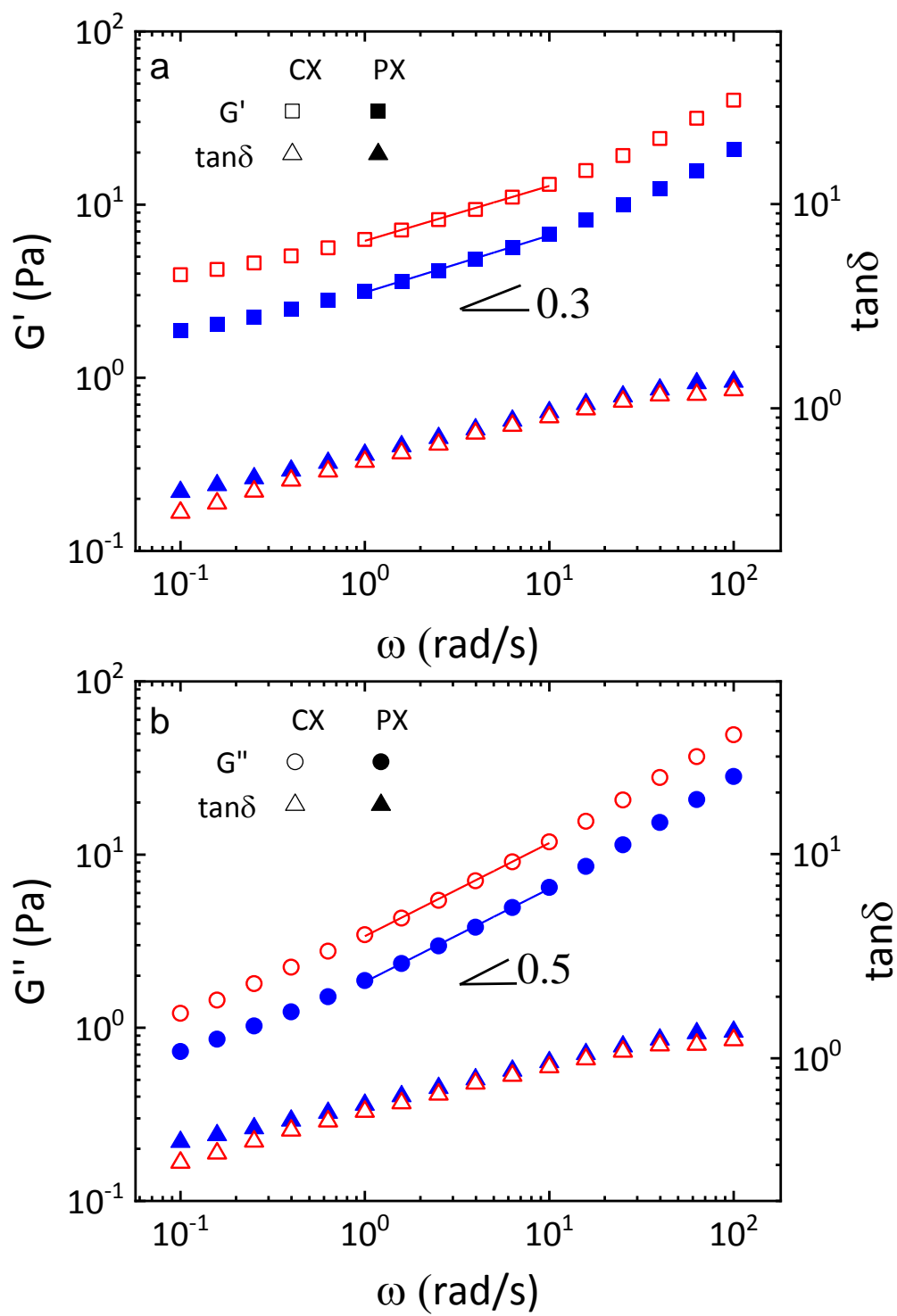

Figure 7 Linear viscoelastic moduli for the CX system at ΔP=0.03 (empty red symbols) and for the PX system at $t_w$=24h (filled blue symbols). (a) Elastic modulus, G' and (b) Viscous modulus, G''. Triangular symbols mark tanδ values.

First, strain sweep experiments for up to 300% strain were performed at several different frequencies (Figure 8). As pointed out above, there is no frequency dependence in the case of the CX system due to the vicinity of the sample to the critical GP. Moreover, as the strain amplitude increases, both non-linear moduli decrease: $G_1'$ values by approximately 20%, and $G_1''$ values only by approximately 5% (cf. Fig. 8a). Thus, the normalized $G_1'$ is always smaller than the normalized $G_1''$ within the range of strains examined. In contrast, for the PX system, an explicit frequency dependence can be observed in both moduli. In addition, the moduli dependence on the strain amplitude differs significantly from that observed for the chemical system. Initially, as the strain amplitude increases the chains between the junction zones align with the deformation field resulting in a drop in $G_1'$ similar in magnitude to that observed for CX. Yet, at a certain strain value, specific to each frequency, a minimum is observed followed by a rise in both moduli values and resulting in normalized $G_1'$ larger than normalized $G_1''$ (cf. Fig 8b). We speculate that the source for this increase is the resistance to deformation by the micro-crystalline junction zones which at much higher strains, will eventually yield manifested by an overshoot, a hint of which is observed for 1 rad/s, a decline in values, and possible breakdown of the network.

Martinetti et al. (2018) and Ramaya et al. (2018) studied PVA hydrogels physically crosslinked by transient complexation with Borax. Their results showed that for the Borax system both moduli increased upon the increase in strain (strain stiffening), a behavior quite different from the one depicted in Figure 8b. Martinetti et al. (2018) suggested that as the deformation increased the crosslinking density increased due to the formation of new Borax-PVA complex sites which lead to the observed rise in both moduli in the non-linear regime.

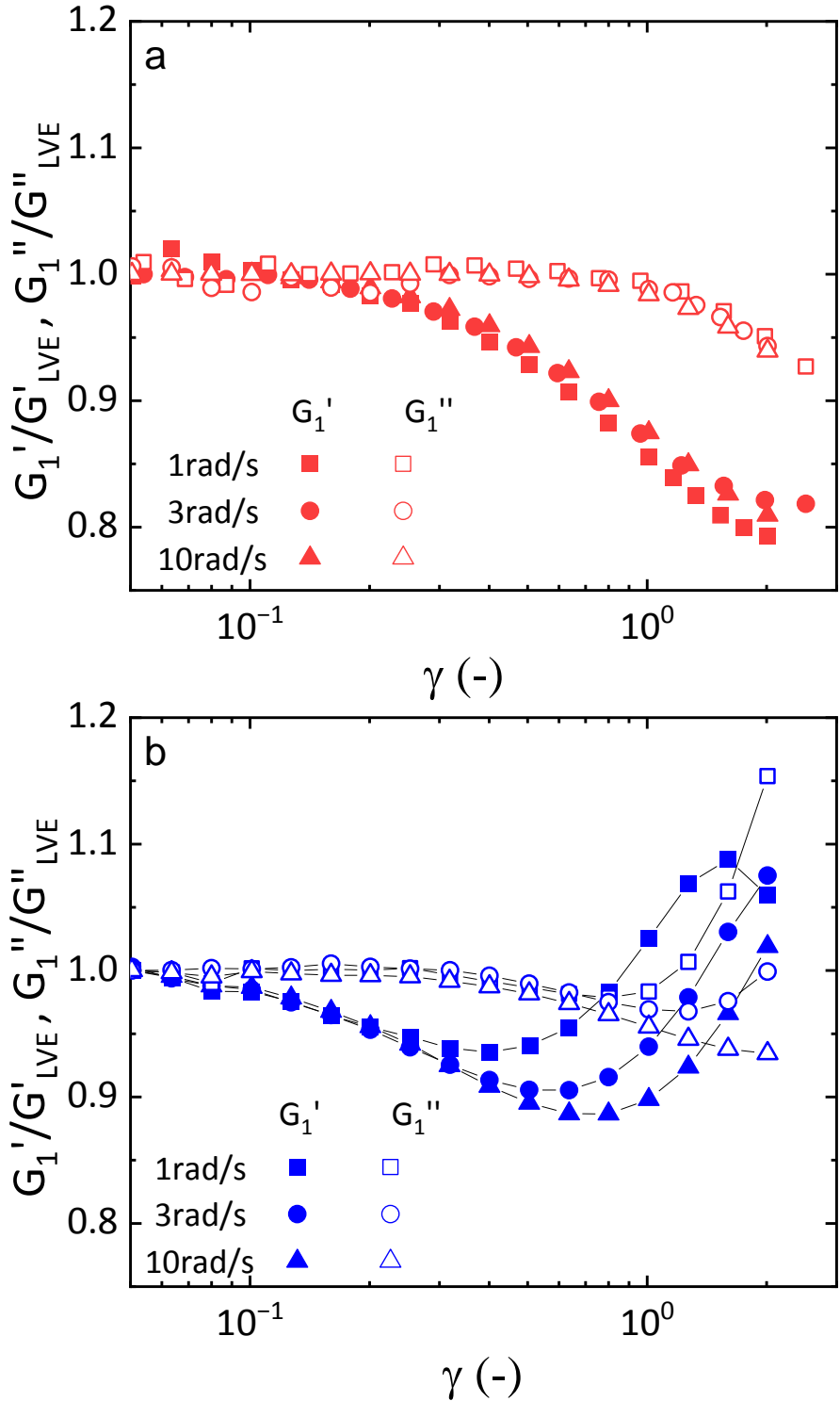

Figure 8 First harmonic non-linear moduli normalized by the moduli in the linear regime as function of strain amplitude, $\gamma_0$, for (a) CX at $\Delta P=0.03$ and (b)PX at $t_w=24h$. filled symbols represents normalized first harmonic non-linear storage modulus, $G'_1/G'_{LVE}$ and open symbols represents normalized first harmonic non-linear loss modulus $G''_1/G''_{LVE}$. The black line is used as a guide for the trend.

From the results presented in Figure 8 it appears that despite the identical characteristics of the polymers (chemical composition, molecular weight, initial concentration) and similarity in SAOS response, the nature and type of crosslinks has a dramatic effect on the non-linear rheology. Furthermore, comparison to the results on Borax crosslinked systems suggests that LAOS experiments also enable differentiation between the two types of physical crosslinking mechanisms.

To investigate the physical meaning of the third harmonic LAOS response the normalized Chebyshev coefficients for both systems were used as depicted in Figure 9. As can be observed in Figure 9a, both $e_3/e_1$ values increase as the strain amplitude increases at all imposed frequencies which means that the material exhibits **local** strain stiffening at large deformations (Sim et al. 2003; Kamkar et al. 2022). **Local** strain stiffening is in line with the concept of alignment of chains with the deformation field. Additionally, the elastic third harmonic response for the PX system begins at lower strain amplitude values and it is considerably larger than that of the CX one in line with the observations related to Fig.8 above.

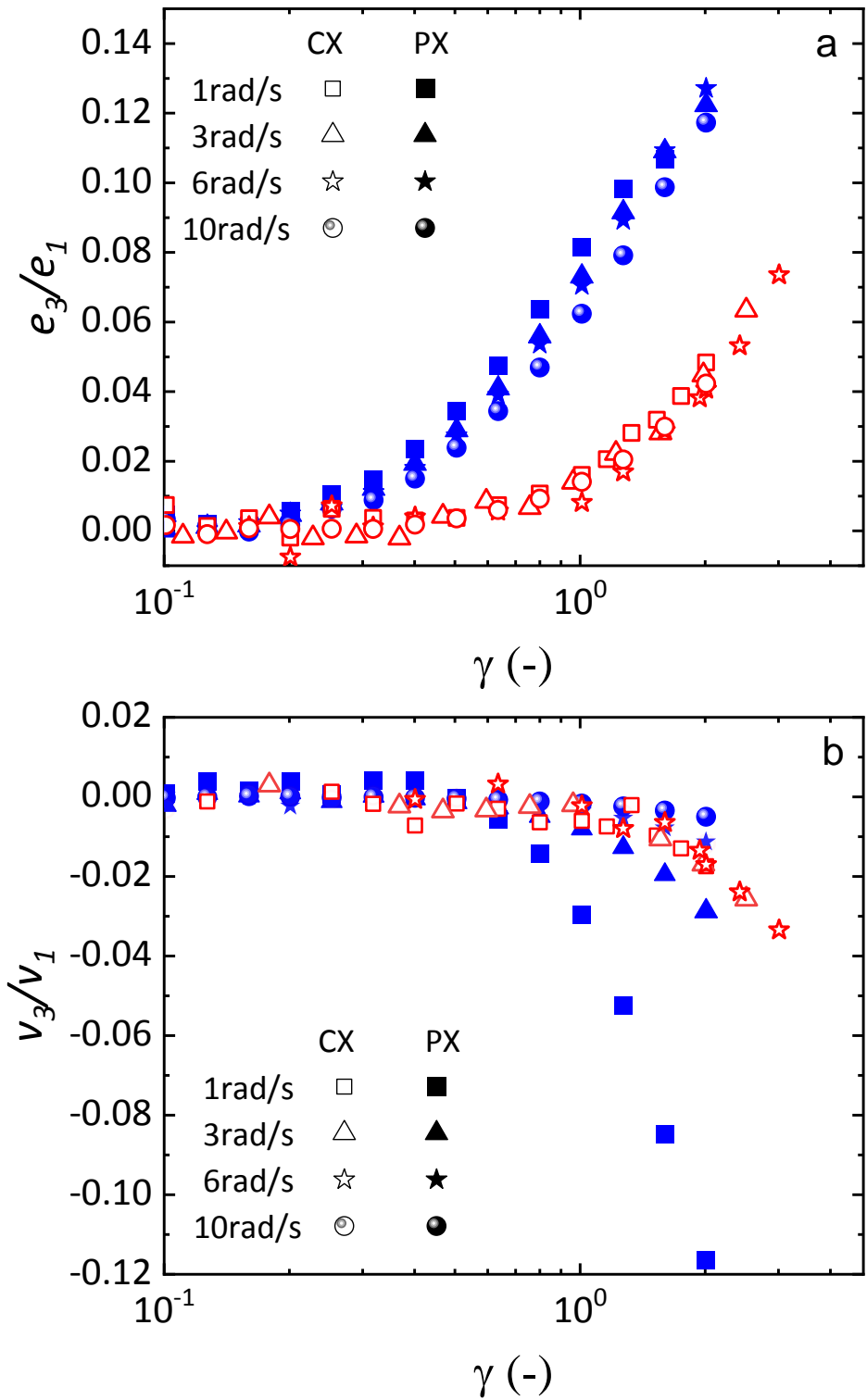

Figure 9 Normalized third Chebyshev coefficients as function of strain amplitude $\gamma_0$, for the CX PVA at $\Delta P=0.03$ systems (empty red symbols) and PX PVA at $t_w=24h$ (filled blue symbols). (a) Normalized elastic Chebyshev, $e_3/e_1$ and (b) Normalized viscous Chebyshev, $v_3/v_1$.

The negligible contribution of the viscous non-linear response to the total shear stress is less than 4% at the highest applied strain in the case of the CX system irrespective of the frequency employed, as evident from the values for the viscous Chebyshev coefficient, $v_3/v_1$, depicted in Figure 9b. However, for the PX system one order of magnitude drop can be observed at the lowest frequency. The large difference in the response of the two systems can be explained as follows: the chemical permanent molecular scaffolding within a network (i.e., covalent crosslinks) do not break during large deformations. In contrast, the reversible crosslinks junction zones based on microcrystalline junctions may disintegrate and at the same time allow the slippage of polymer chains or segments which were trapped during the formation of the micro-crystals (Hyun et al. 2002; Danielsen et al. 2021). Since experiments are carried out above the microphase dissolution temperature no new domains are formed in contrast with Borax complexation mechanism in which new crosslinks are being formed and reformed upon deformation (Martinetti et al. 2018).

Meaningful interpretation of LAOS nonlinearities can be effectively represented by the Lissajous-Bowditch (in short, Lissajous) plots (Ewoldt et al. 2008a; Hyun et al. 2011). In the linear viscoelastic regime both elastic (strain) and viscous (strain rate) Lissajous curves are represented by perfectly symmetrical ellipses. The ellipsoidal shape is the result of the fundamental sinusoidal nature of both stress and strain in the linear regime devoid of any higher harmonic contributions to the stress response. Any distortion from the elliptical shape marks the presence of higher harmonics in the stress response and the onset of non-linear behavior (Goudoulas and Germann 2019).

Figure 10 illustrates the elastic and the viscous Lissajous curves for different strain amplitudes at $\omega=1$rad/s. Elastic Lissajous curves for the CX at $\Delta P=0.03$ and for PX at $t_w=24$h are presented in Figures 10a and 10b, respectively while the viscous Lissajous curves are presented in Figures 10c and 10d. The Lissajous representation is easily related to the third Chebyshev coefficients since their values determine the direction of the skewed distortion of the ellipsoidal curves (Ewoldt and Bharadwaj 2013). In Figure 10b, we note that the elastic Lissajous curves at the highest strain amplitude are skewed in the counterclockwise direction for the PX system, suggesting the $e_3 > 0$ i.e., the system exhibits local strain stiffening. However, no skewedness or distortion is observed in Figure 10a for the CX system at the same applied strain amplitude. Referring to Figure 9a we note that at 200% strain the 3rd elastic Chebyshev coefficient for PX PVA is three times larger than that for CX PVA and as a result the non-linearity in the latter is too small to yield detectable skewness of the ellipse.

The viscous Lissajous curve for the PX system (Figure 10d) at highest strain amplitude is slightly skewed in the clockwise direction suggesting that $v_3 < 0$ which means that the material exhibits local strain rate thinning (Ewoldt and Bharadwaj 2013). Yet, at the same strain amplitude no distortion of the ellipsoids is observed for the CX system (Figure 10c) implying no discernible non-linear response in agreement with Figure 9b in which PX exhibits 12 times larger $v_3$ value than CX. The Lissajous curves clearly demonstrate the large difference in the non-linear behavior between the two systems examined. Since the molecular and linear rheological characteristics of the two systems are identical, these differences are clearly the result of the differences in the crosslinks holding the hydrogels together.

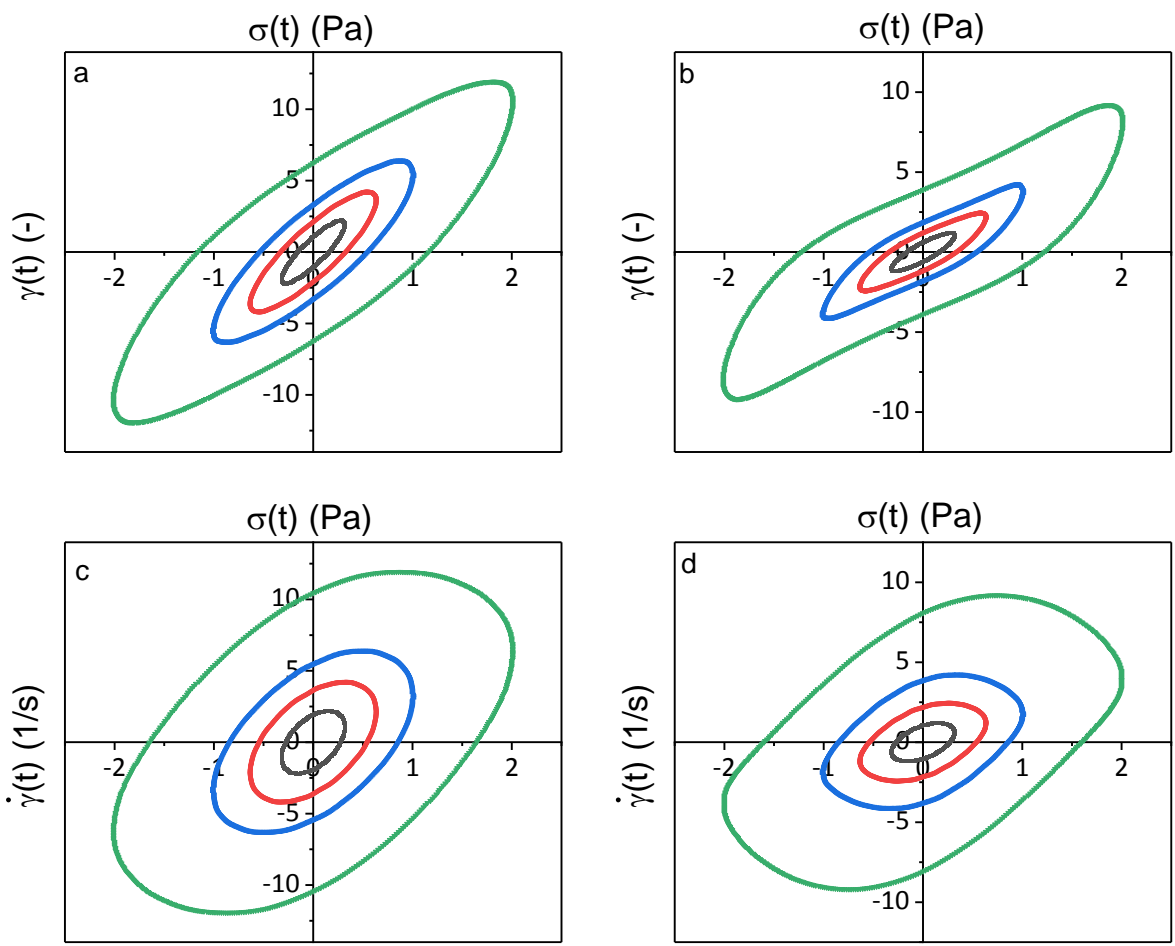

Figure 10 The elastic and viscous Lissajous curves at ω=1 rad/s and different strain amplitudes (30%, 60%, 100%, 200%) for the chemically crosslinked PVA at ΔP =0.03 and the physically crosslinked PVA $t_w$=24h. Elastic Lissajous curves for (a) CX system and (b) PX system. Viscous Lissajous curves for (c) CX system and (d) PX system.

**Comparison between CX and PX systems in fully developed hydrogels**

Now we turn our attention to two systems in the post-gel state which exhibit similar rheological characteristics in the linear regime: a CX hydrogel at Δ$P$=0.1 and a PX hydrogel after four f/t cycles ($t_w$=24h at each cycle). From the data in Figure 11 it is evident that the polymeric networks are relatively far from the critical GP since $G'$ values for both systems are independent of the imposed frequency over two decades of frequency and are substantially larger than $G''$ which also shows only a weak dependence on frequency. In addition, the values of the phase angle, $\delta$, are very low.

The analysis of the non-linear rheological behavior of these two systems in terms of the first harmonic complex shear moduli, the Chebyshev coefficients, and the corresponding Lissajous curves are presented in Figures 12, 13, and 14 respectively.

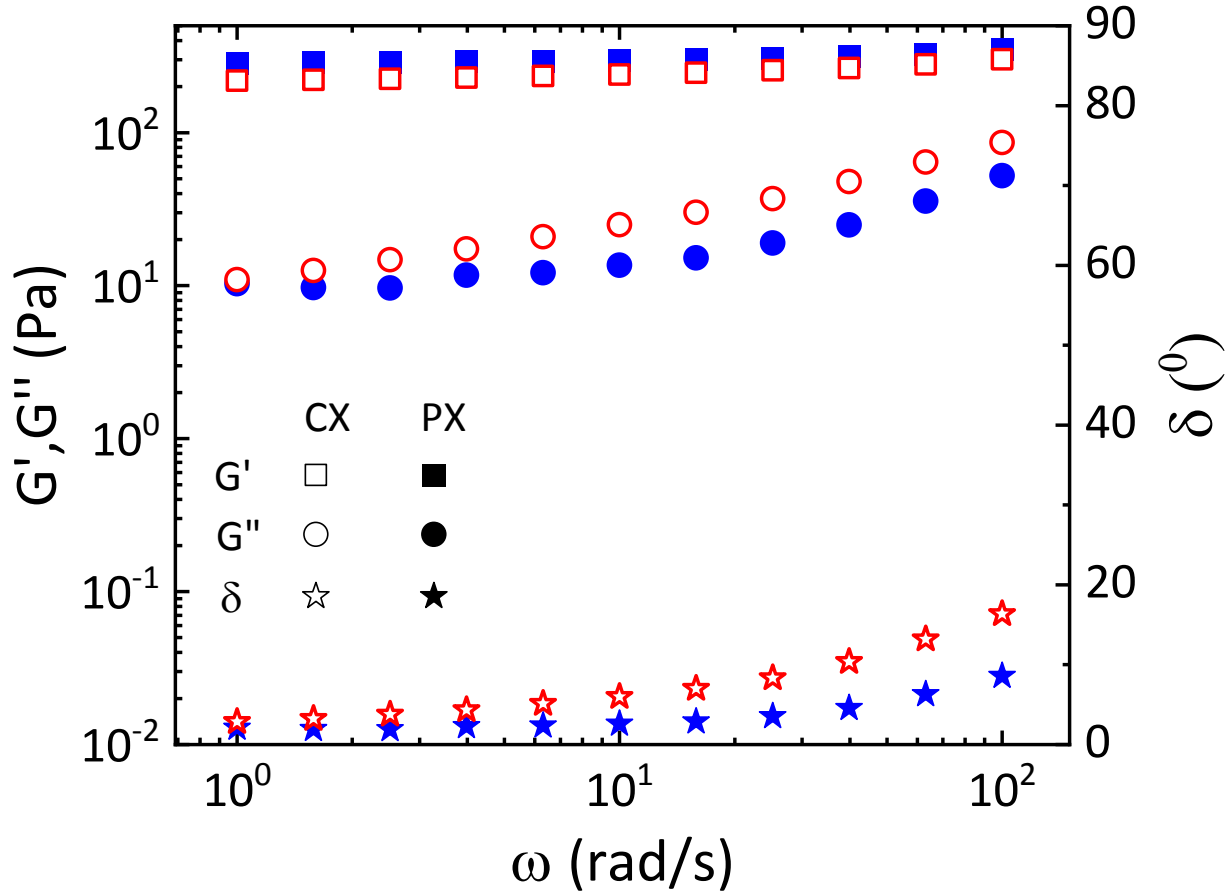

Figure 11 Linear viscoelastic moduli for the CX PVA at ΔP=0.1 (empty red symbols) and PX PVA after the 4[th] freeze/thaw cycle (filled blue symbols).

As illustrated in Figure 12a, over the range of imposed strain amplitudes, CX exhibited only a small (~ 10%) reduction in $G_1^{'}$ values contrasted with one order of magnitude reduction for the PX system. In addition, the onset of $G_1^{'}$ decline is delayed till strain values one order of magnitude larger for the CX system. The non-linear characteristics of $G_1^{''}$ for the two systems depicted in Figure 12b, differ in magnitude, shape, and onset. For both systems the 1[st] harmonic loss modulus increases with increasing strain, but in the case of PX the increase is followed by a sharp decrease resulting in a pronounced overshoot. Since the onset of upturn is delayed by one order of strain magnitude for the CX system so is the overshoot (cf. Figure 5d). In addition, as with the storage modulus, the magnitude of the overshoot is up to one order of magnitude larger for PX. Based on these results, it is evident that while the mechanical properties in the linear regime are quite similar, the non-linear rheological response of the PX system is significantly enhanced relative to the CX system. While there are no major differences between the non-linear effects in the elastic modulus of CX systems at $\Delta P= +0.03$ (Figures 5c and 8a) and $\Delta P=+0.1$ (Figures 5d and 12a) as already discussed, for the PX system a drastic increase in the non-linear response is observed between the 1[st] f/t cycle (Figure 8) and the 4[th] f/t cycle (Figure 12). Ricciardi et al. (2004a, b) investigated the degree of crystallinity of PVA hydrogels formed by consecutive freeze/thaw cycles. They determined that a small number of cycles resulted in small and highly hydrated crystals with relatively large amounts of imperfections. Increase in the number of cycles resulted in larger and less hydrated crystals. Since these crystals serve as the junction zones in the PX hydrogels we believe that the increase in the non-linear response is due to the change in the strength and structure and the reinforcement afforded by the crystalline crosslinking junctions.

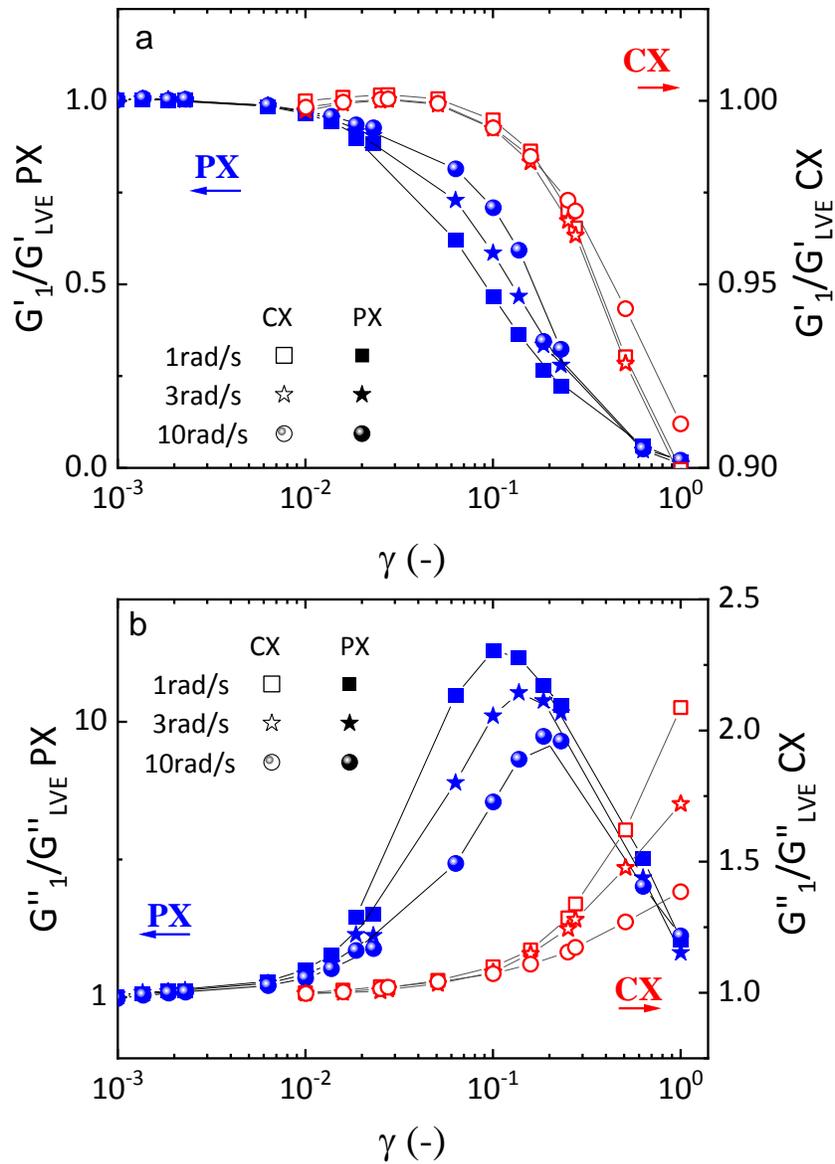

Figure 12 First harmonic complex shear moduli normalized by the moduli values in the linear regime as function of strain amplitude, $\gamma_0$, for the CX hydrogel at $\Delta P=0.1$ (open red symbols) and the PX hydrogel after the 4th freeze/thaw cycle (filled blue symbols). (a) Normalized first harmonic non-linear storage modulus, $G'_1/G'_{LVE}$ and (b) normalized first harmonic non-linear loss modulus $G''_1/G''_{LVE}$. The lines serve as guides for the trend.

The large difference between the PX and CX systems is further magnified while examining the third Chebyshev coefficients, $e_3$ and $v_3$, and the corresponding Lissajous curves. The magnitude of the normalized elastic Chebyshev coefficient, $e_3/e_1$ for CX is negligible as illustrated in Figure 13a. In contrast, for the PX system at the lowest applied frequency (1 rad/s), this contribution is approximately 80% strain stiffening ($e_3 > 0$). Surprisingly, $e_3/e_1$ for 10 rad/s is considerably smaller than that for 1 rad/s in contrast with the small dependence on frequency in the normalized $G_1^{'}$ (Fig. 12a). The normalized viscous Chebyshev coefficient, $v_3/v_1$, (Figure 13b) similarly reveals the absence of any noticeable nonlinear third harmonic contribution for the CX system. As discussed previously the system is relatively far from the GP, and therefore, comprises more developed network with less defects (dangling chains, unattached clusters) which explain the lack of non-linear viscous contribution. A somewhat larger contribution with what appears to be a flat extended overshoot pattern is observed for the PX system.

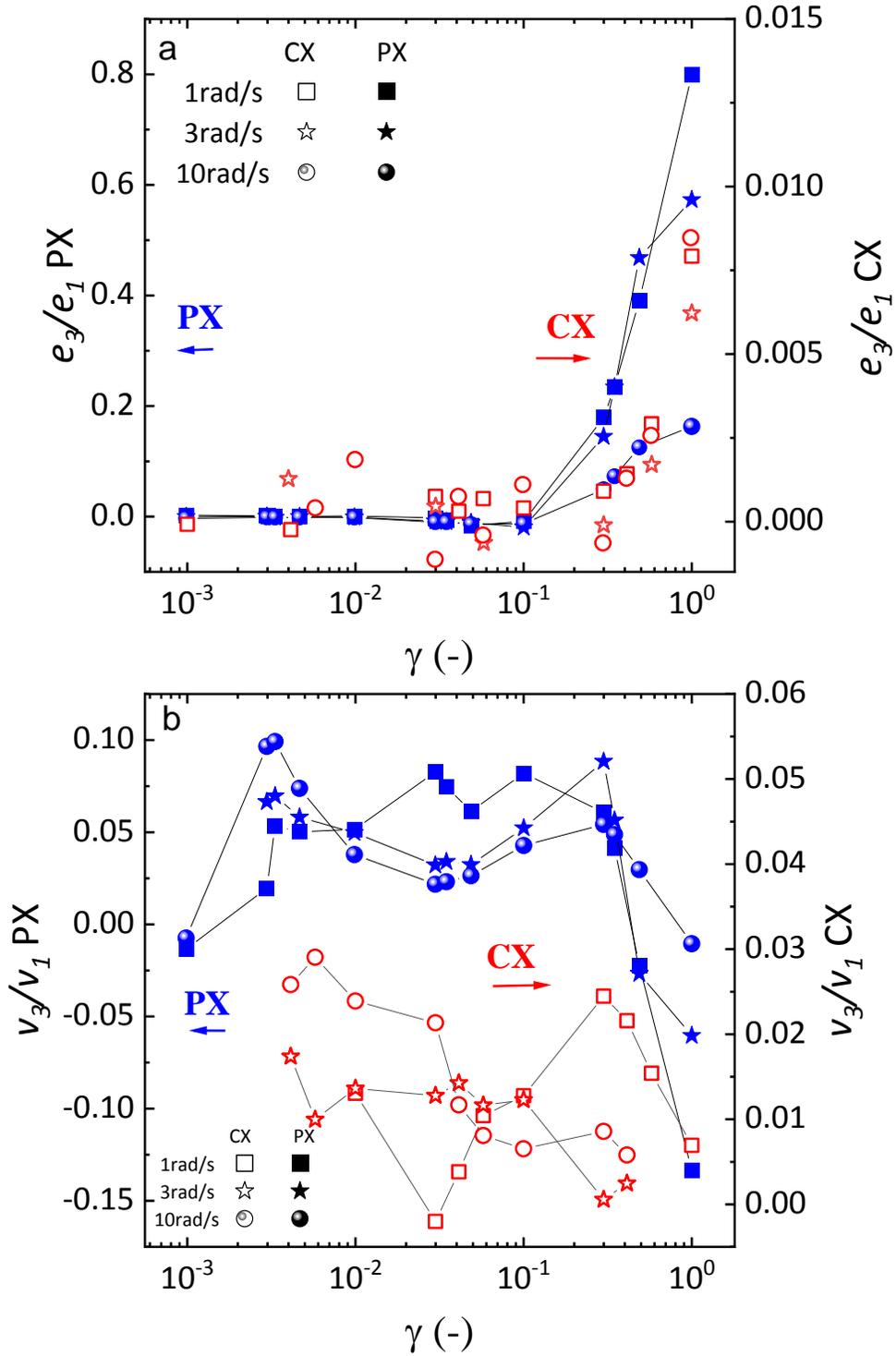

Figure 13 Normalized third Chebyshev coefficients as function of strain amplitude $\gamma_0$, for the CX hydrogel at $\Delta P=0.1$ (empty red symbols) and the PX hydrogel after the 4th freeze/thaw cycle (filled blue symbols). (a) Normalized elastic Chebyshev, $e_3/e_1$ and (b) normalized viscous Chebyshev $v_3/v_1$.

The Lissajous curves of the chemically crosslinked system, Figure 14a and 14c show a linear response within the imposed strain amplitude range in agreement with the relatively small changes in the first harmonic moduli (Fig. 12) and third harmonic Chebyshev coefficients (Fig. 13). As a result of the strong non-linear elastic contribution of $e_3$ in the case of the PX system the shape of the Lissajous curves, in Figure 14b and 14d, exhibit a peculiar tilted and highly distorted ellipsoid shape especially in the highest strain. At 50% and even more so at 100% strain for extended range of shear values the stress is almost independent of strain and gives rise to self-intersection and secondary loops in the shear rate. The peculiarity of these highly distorted Lissajous has been addressed by Rogers and Letinga (2012) and Paulos et al. (2013). The latter attributed these features to phenomena such as plastic flow and yielding (cf. discussion in Paolos et al. (2013) in relation to Fig. 4 thereof). It should be pointed out that unlike all other systems examined here, for this particular system fifth order harmonics are probably not negligible which could be the source of these observations. Yet, the large differences between the nicely ordered Lissajous curves for the CX system (Fig 14 a and c) and those for the PX system (Fig. 14 b and d) serve in a remarkable way the purpose of distinguishing between the two types of crosslinks.

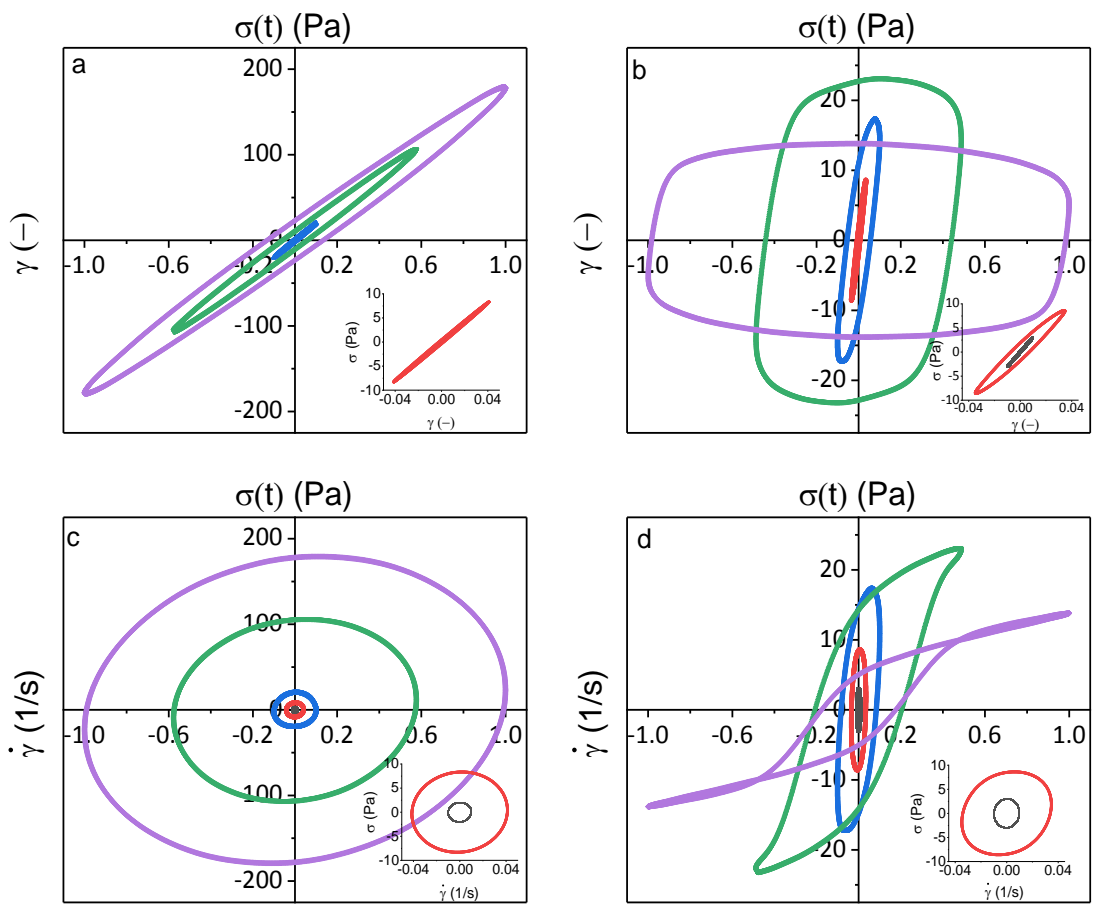

Figure 14 Elastic and viscous Lissajous curves at ω=1 rad/s for the CX hydrogel at ΔP =0. and PX hydrogel after the 4th freeze/thaw cycle systems at different strain amplitudes (1%, 5%, 10%, 50%, 100%). Elastic Lissajous curves for (a) CX system and (b) PX system. Viscous Lissajous curves for (c) CX system and (d) PX system. Inset curves represent the linear response at small strain amplitudes.

The results provided above further emphasize the findings that despite both systems exhibiting comparable mechanical properties within the linear regime, they significantly diverge in their non-linear responses. The utilization of strain sweep, Chebyshev coefficients obtained through Fourier transformation, and Lissajous analysis proves to be highly effective in differentiating between these systems and even between different types of physical crosslinks. These findings imply that the distinctive non-linear rheological behavior is intricately linked to the nature of the crosslinks. Nonetheless, the underlying structural factors causing such contrasting non-linear responses require further elucidation by means of complementary experimental techniques such as rheo-NMR and the like.

## Conclusions

In the present work, we investigated the SAOS, and LAOS behavior of two PVA crosslinked systems differing by their crosslinking mechanism – chemically vulcanized crosslinked system (CX) and physically crosslinked system by the freeze/thaw process (PX).

Initially, it was found for the CX system that the value of the critical exponent $n$ determined by the W-C analysis, was identical to the one obtained by the hyper-scaling relation affirming the robustness and validity of the results. For the PX PVA system, the quasi-state equilibrium moduli values were measured using the inverse-quenching freeze/thaw technique (Hassan and Peppas 2000; Avallone et al. 2021). Using this technique, the critical GP was determined, and the corresponding critical relaxation exponent, $n$, gel stiffness, $S$, and longest relaxation time, $\tau_{max}$ values were determined for both systems.

The non-linear response of the CX system was studied by means of strain sweep experiments at different frequencies for different relative distances from the GP, $\Delta P$. From the results, it was found that in the vicinity of the GP, the frequency dependence vanished, in contrast to systems far from the GP. This result suggests that the dynamic self-similarity phenomena are preserved in the non-linear regime.

Using the non-linear results of the strain sweep experiments at strain amplitude, $\gamma_0 = 200\%$, at different frequencies, for the PX and CX systems, W-C curves were constructed. The location of the critical degree of crosslinking, $P_c$, obtained in the non-linear regime was identical to the one obtained by SAOS experiments. The calculated critical exponent $n$ and the corresponding fractal dimension, $d_f$, for both systems were similar to the ones in the linear regime. However, the gel stiffness, $S$, was lower by approximately 15% for the CX system and 40% for the PX system. It appears that the W-C method may be extended to apply in the non-linear regime. To test the universality of this finding, examination of a large number of different crosslinked systems is required (Kogan and Gottlieb 2025).

Two systems relatively close to the GP, one CX and one PX, with similar linear rheological characteristics were compared by means of LAOS experiments and analyzed in terms of strain sweep, Chebyshev coefficients, and Lissajous curves. A remarkably different non-linear behavior was observed in the moduli-strain amplitude dependence. In addition, the third elastic non-linear harmonic response of the PX system was found to be considerably larger than that of the CX system. These results clearly indicated that the elastic non-linear rheological response is quite sensitive to the type of crosslinking holding the hydrogels together. Apart from the viscous third harmonic response of PX at the lowest frequency tested, only minor contribution of the viscous response is observed for systems in the vicinity of the gel point.

Finally, comparing two crosslinked systems relatively far from the GP with identical linear rheological properties revealed that here as well large differences exist between the two network types. The non-linear response of the physically crosslinked system increased dramatically whereas, only negligible non-linear response was demonstrated by the CX system. The results presented in this work clearly indicate that the nature of the crosslinks strongly affects the non-linear rheological response. It appears that LAOS rheology is a powerful tool for studying polymeric networks and revealing additional data unattainable by linear rheology.

**Appendix 1: Stress relaxation experimental results for chemically and physically crosslinked PVA hydrogels.**

The applied strain was varied between samples in the range 0.1% and 20% all in the linear viscoelastic regime. The equilibrium modulus was obtained by extrapolation to infinitely long time by means of cubic spline as illustrated by the dashed line for the upper curve in Figure 15a. The piecewise cubic spline interpolation/extrapolation procedure employed is the one available from the Origin graphic software. The general equation for the cubic polynomial within a segment is $S_i(x) = a_i x^3 + b_i x^2 + c_i x + d_i$, where *i* represents the segment number. The coefficients ($a_i$, $b_i$, $c_i$, $d_i$) are determined to ensure the spline passes through the given data points and has continuous first and second derivatives at the boundaries between segments. These cubic polynomials are used to determine the $G_e$ value by extrapolation to the asymptote at long times.

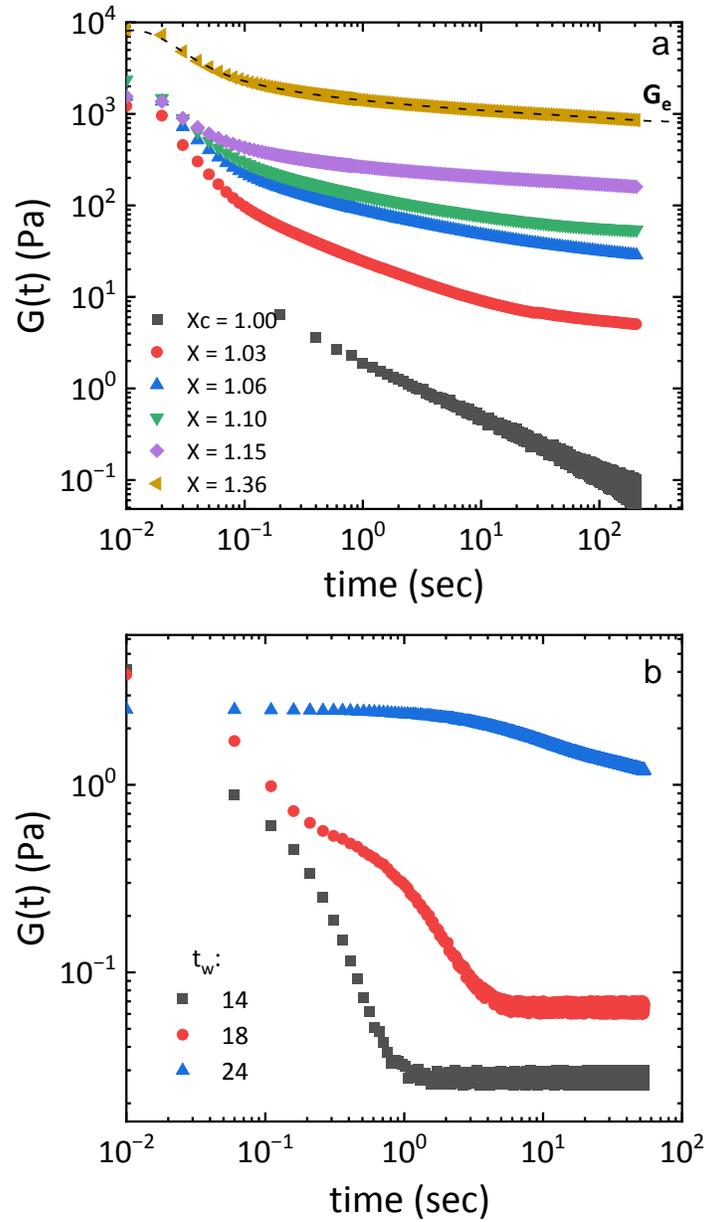

Figure 15 Relaxation modulus G(t) for (a) different degree of crosslinking X for chemically crosslinked PVA and (b) different degree of crosslinking $t_w$ for physically crosslinked PVA

**Appendix 2: Procedure used for the estimation of the zero-shear viscosity for the pregel physically crosslinked PVA hydrogels.**

The modified Cross empirical model was used to fit the complex viscosity data obtained by frequency sweep experiments on pregel systems. Slope values (n) were obtained directly from the $\eta^*(\omega)$ data, $\eta_\infty$ was set to zero, leaving two parameters to be determined: m and $\eta_0$. Four sets of data are available: 1) solution of uncrosslinked PVA (12%); 2) $t_w$=2 hours (pregel) 3) $t_w$= 4 hours (pre-gel) 4) $t_w$=8 hours (gel point). The following constraints are defined:

$$\eta_{0,sol} < \eta_{0,2hr}$$
$$\eta_{0,2hr} < \eta_{0,4hr}$$
$$m_{sol} < m_{2hr}$$
$$m_{2hr} < m_{4hr}$$

By an iterative process the difference between the two pregel normalized curves in Fig. 4b was minimized.

**Acknowledgements**

D.K. acknowledges the financial support and travel grants provided by the BGU Chemical Engineering Graduate Research Program.

**Conflict of interest**

The authors have no conflicts to disclose.